%% file: PRB_DoubleProvskite.tex
\documentclass[aps,pra,reprint,superscriptaddress,longbibliography]{revtex4-1}
\usepackage{graphicx,color}
\usepackage{amsmath, amssymb}
\usepackage{longtable}
\usepackage{natbib}
\usepackage{bm}
\usepackage[hidelinks]{hyperref}
\hypersetup{
  colorlinks=true, 
  urlcolor     = blue, 
  linkcolor    = blue, 
  citecolor   = red 
}

\usepackage{multirow}

\usepackage{graphics,amssymb,amsmath,epsfig,color}
\usepackage{graphicx}

\usepackage{rotating}

\begin{document}

\title{First-principles derivation of elastic interaction between Jahn-Teller centers in crystals via lattice Green's functions}

\author{Zhishuo Huang}
\email[]{zhishuohuang@gmail.com}
\affiliation{Department of Chemistry, National University of Singapore, Block S8 Level 3, 3 Science Drive 3, Singapore 117543, Singapore}
\author{Naoya Iwahara}
\affiliation{Graduate School of Engineering, Chiba University, 1-33 Yayoi-cho, Inage-ku, Chiba 263-8522, Japan}
\author{Liviu F. Chibotaru}
\email[]{liviu.chibotaru@kuleuven.be}
\affiliation{Theory of Nanomaterials Group, KU Leuven, Celestijnenlaan 200F, B-3001 Leuven, Belgium}
\date{\today}

\begin{abstract}
Jahn-Teller (JT) systems with strong and intermediate vibronic coupling are described in terms of local JT active vibrational modes. 
In JT crystals, the elastic interaction of these modes at different JT centers plays a crucial role, for instance, in determining critical temperature of structural phase transitions. 
Despite their importance, the parameters of elastic interaction between JT centers have not been accessed yet by first-principles calculations. 
In this paper, we develop an effective Hamiltonian methodology for the thorough description of the elastic interactions in cooperative Jahn-Teller problems, which treats the interactions with both phonons and uniform strains. 
All the microscopic parameters, such as lattice Green's functions, elastic modulus,  can be obtained or calculated base on first-principles the calculation. 
The method has been applied to a series of 5$d^1$ double perovskites. 
Such effective Hamiltonian methodology can be, in general, used to investigate the APES of any type of the local distortions, such as impurities, defects, etc.
\end{abstract}

\maketitle
\section{Introduction}
Multi-center vibronic interactions are found in a wide number of systems such as impurity Jahn-Teller (JT) centers\cite{Perlin1984TheDJ}, multi-center molecular clusters \cite{Willett1985magneto} and JT crystals \cite{Riste1977electron,Bersuker1989VibronicII}. In these compounds, except for the $E \otimes b_1$, $E \otimes b_2$, and $T \otimes e$ the vibronic problem, where the vibronic interaction on the centers does not mix different electronic functions, the vibronic problem becomes rather involved, because the localized electrons in degenerated orbital states interact with the JT active distortions of the JT centers, while these JT centers interact via the common vibrational modes (phonons) of the system \cite{HZSGehring1975co-operative, Kaplan2012cooperative}.

Regarding the co-operative JT effect, an one-center nuclear coordinates based approach was proposed for the investigation of the adiabatic potential of molecular systems\cite{Chibotaru1992onecenter, Chibotaru1994adiabatic}. However, such an approach seems to be less adequate in the case of two interacting Jahn-Teller impurities in the crystal, in which the derivation of the model Hamiltonian for the static JT effect \cite{Bersuker1990effects} lack of substantiation. Another method, assuming that the displacement fields due to JT impurities are independent, has been developed, as a perfect approximation for the system where the JT impurities are sufficiently distant \cite{Novak1969interactions, Novak1970some}. 
Thus, the theory of vibronic interactions in these systems is not well-developed compared with one-center systems.

In this paper, with the microscopic parameters from first-principles calculations, an effective Hamiltonian for a static cooperative JT effect is proposed. This methodology is then applied to a series of double perovskites (DP) A$_2$B'BX$_6$.

B'-site ordered double perovskites, of which the structure is shown in Fig. \ref{Fig:OctahedralConf} (a)., have attracted increasing interests because of its potential to host diverse exotic states of matter\cite{Witczak2014correlated}. Different combination of magnetic or nonmagnetic cations of A, B' and B sites will lead to different magnetic phases\cite{Vasala2015perovskites}. There are three series of these DP materials with B electronic configurations n$d^1$ ($T_{2g}^1$), n$d^2$ ($T_{2g}^2$), and n$d^3$ ($T_{2g}^3$), of which each series presents different issues. For example, the crystallographic distortions of $T_{2g}^1$ and $T_{2g}^2$ ions are potentially attributed to JT effect, while $T_{2g}^3$ ions are not. In addition, the orbital moment is not fully quenched in $T_{2g}^1$ and $T_{2g}^2$ ions while n$d^3$ ions are orbital singlets. Recent publications summarize in tabular form what is known about the $T_{2g}^2$ and $T_{2g}^3$ cases\cite{Thompson2014long, Thompson2016frustrated, Yuan2015high}. For the simplicity, 5$d^1$ ($T_{2g}^1$) double perovskites will be the targeted materials for the application. 

\section{Effective Hamiltonian for cooperative Jahn-Teller effect}
The basic mechanism for the JT effect is the interaction between nuclei vibrations and the electronic states. 
For isolated JT active centers, the methodology is clear as discussed by plenty of articles \cite{Liu2018dynamical, Liu2018quadratic, Huang2020dynamical, Bersuker1989VibronicII}. 
When the JT centers are embedded in crystals, the properties of nuclei vibrations and the electronic structure will be altered by the collective vibrations in the crystal and the symmetry of crystal field, respectively. 
The properties of one JT active centers may be influenced by the neighboring JT active centers or even the JT active centers far way via phonons. 
Therefore in the study of cooperative JT effect, the vibrational part should be substituted by the analysis of phonon properties. 
Specially, when the JT active centers presenting in a crystal lattice have a sufficiently high concentration, the stability of the entire crystal can be destroyed, which leads to the uniform strains. Thus, the thorough understanding of cooperative JT effect in crystals should consist of the investigation of the phonon contribution (elastic interaction part) and the uniform strain contribution \cite{HZSGehring1975co-operative, Kaplan2012cooperative}. 

\subsection{Contribution from phonons}
\label{subsec:phononContribution}
The potential energy of the Jahn-Teller crystal is described by the operator
\begin{equation}
\label{eq:JToprator_phonon}
\hat{H}=\sum_{\mathbf{k}j}\frac{1}{2}\omega_{\mathbf{k}j}^{2}Q_{\mathbf{k}j}Q_{\mathbf{k}j}^{*}+\sum_{\mathbf{n}}\hat{U}_{vib}^{\mathbf{n}}(\{q_{\mu\mathop{\Gamma\gamma}}^{\mathbf{n}}\})
\end{equation}
where $\mathbf{k}$ and j denote the wave vector and phonon branch, respectively, $Q_{\mathbf{k}j}$ is the phonon modes and $\omega_{\mathbf{k}j}$ the corresponding frequency. $\hat{U}_{vib}^{\mathbf{n}}$, expressed on the basis of a set of (pseudo)degenerate electronic functions localizing at the corresponding center, describes the vibronic interaction with local nuclear distortions, donated by $q_{\mu\Gamma\gamma}^{\mathbf{n}}$ , which transforms after irreducible representation $\Gamma$ of the site symmetry group at the $\mathbf{n}$th site of the crystal lattice.

The adiabatic potential energy surface (APES) can be obtained by the summing up the diagonalized potential energy operator ($\hat{U}_{vib}^{\mathbf{n}}$) on each JT active centers \cite{Englman1972jahn}, expressing as 
\begin{equation}
\label{eq:APES_total}
U_{\alpha_{1}\alpha_{2}...}=\sum_{\mathbf{k}j}\frac{1}{2}\omega_{\mathbf{k}j}^{2}Q_{\mathbf{k}j}Q_{\mathbf{k}j}^{*}+\sum_{\mathbf{n}}\epsilon_{\alpha_{\mathbf{n}}}^{\mathbf{n}}(\{q_{\mu\Gamma\gamma}^{\mathbf{n}}\})
\end{equation} 
where $\alpha_{\mathbf{n}}$ numbers the eigenvalues of the vibronic operator at the $\mathbf{n}$th center. 

Since the phonon modes ${Q_{\mathbf{k}j}}$ form a complete set basis for the lattice dynamics, as a complete set basis for the dynamics of local JT active centers, local JT active modes $\{q_{\mu\Gamma\gamma}^{\mathbf{n}}\}$ can be expressed by the the phonon modes by Van Vleck coefficient $a_{\mu\Gamma\gamma}^{\mathbf{n}}(\mathbf{k}j)$ \cite{Van1940paramagnetic} as:
\begin{equation}
\label{eq:VanVleckexpansion}
q_{\mu\Gamma\gamma}^{\mathbf{n}}=\sum_{\mathbf{k}j}a_{\mu\Gamma\gamma}^{\mathbf{n}}(\mathbf{k}j)Q_{\mathbf{k}j}.
\end{equation}

The APES extrema can usually be found by Lagrange multiplier. However, this method can only be applied straightforward for simple JT problem, such as $E\otimes b_1$, $E\otimes b_2$, $T\otimes e$, because in these systems, the non-adiabaticity is no longer important, resulting in the only shifts of equilibrium phonon coordinates. While when the interaction between different JT centers are not negligible, the eigen values $\epsilon_{\alpha_{\mathbf{n}}}^{\mathbf{n}}$ are complicated functions of $Q_{\mathbf{k}j}$, which makes the extrema equations unsolvable for systems with a large number of vibrational modes. Good thing comes when comparing the applications of Lagrange multiplier\cite{Korn2000mathematical} to both $\{Q_{\mathbf{k}j}\}$, $\{q_{\mu\Gamma\gamma}^{\mathbf{n}}\}$ based on the condition defined in Eq. \ref{eq:VanVleckexpansion}. Due to the parity symmetry ($Q_{\mathbf{k}j}=Q^{*}_{-\mathbf{k}j}$) \cite{Maradudin1963theory}, the complex phonon coordinates will not double the number of corresponding extrema equations from the Lagrange multiplier. Following the procedure in Appendix \ref{sec:appendix_effectiveElastic}, we have reduced the system of extrema equations for the sheet $U_{\alpha_{1}\alpha_{2}...}$ to a limited number of equations depending on one-center JT active modes: 
\begin{equation}
\label{eq:Extrema_expre}
\frac{\partial\epsilon_{\alpha_{\mathbf{n}}}^{\mathbf{n}}(\{q_{\mu\Gamma\gamma}^{\mathbf{n}}\})}{\partial q_{\mu\Gamma\gamma}}+\sum_{\mathbf{n}'\mu'\Gamma'\gamma'}K_{\mathbf{n}'\mu'\Gamma'\gamma'}^{\mathbf{n} \mu\Gamma\gamma}q_{\mu'\Gamma'\gamma'}^{\mathbf{n}'} = 0
\end{equation}
with 
\begin{equation}
\label{eq:KXirelation1} 
K=\Xi^{-1} 
\end{equation}
\begin{equation}
\label{eq:expre_zeta_main}
\Xi_{\mathbf{n}'\mu'\Gamma'\gamma'}^{\mathbf{n}\mu\Gamma\gamma} = \sum_{\mathbf{k}j}\frac{a_{\mu'\Gamma'\gamma'}^{\mathbf{n}'*}(\mathbf{k}j)a_{\mu\Gamma\gamma}^{\mathbf{n}}(\mathbf{k}j)}{\omega^2_{\mathbf{k}j}}.
\end{equation}

Detailed analysis of Eq. \ref{eq:KXirelation1} and Eq. \ref{eq:expre_zeta_main} can be found as following.

\subsubsection{Derivation of the effective force matrix }
\label{sec:appendix_effective}
From the derivation of the effective Hamiltonian in Appendix. \ref{sec:appendix_effectiveElastic}, the matrix $\Xi$ and $K$ are connected via $K=\Xi^{-1}$. 
Due to the periodicity, all the relevant property should be treated in reciprocal space\cite{Martin2004electronic}, though both $K$ and $\Xi$ describe the elastic property in real space.
To distinguish the parameters in both reciprocal space and real space, we write $\Xi^{\mathbf{n}\mu \Gamma \gamma}_{\mathbf{n}'\mu' \Gamma' \gamma'} $, $K^{\mathbf{n}\mu \Gamma \gamma}_{\mathbf{n}'\mu' \Gamma' \gamma'}$ as $\Xi^{\mu \Gamma \gamma}_{\mu' \Gamma' \gamma'} (\bm{n})$, $\Xi^{\mu \Gamma \gamma}_{\mu' \Gamma' \gamma'} (\bm{n})$, because they are parameters depending on the distance between different JT active centers, while the corresponding matrix elements in reciprocal space are denoted as $\Xi^{\mu \Gamma \gamma}_{\mu' \Gamma' \gamma'} (\bm{k})$, $\Xi^{\mu \Gamma \gamma}_{\mu' \Gamma' \gamma'} (\bm{k})$.

First, the $\Xi^{\mu \Gamma \gamma}_{\mu' \Gamma' \gamma'} (\bm{n})$ should be transformed into its representation in reciprocal space,
\begin{equation}
\Xi ^{\mu \Gamma \gamma}_{\mu' \Gamma' \gamma'} (\bm{k})=\sum_{\mathbf{n}} \Xi ^{\mu \Gamma \gamma}_{\mu' \Gamma' \gamma'} (\mathbf{n}) exp({i\bm{k} \cdot \mathbf{n}}).
\end{equation}

Second, for each $\bm{k}$, the inverse matrix of $\Xi (\bm{k})$ is calculated, which is the elastic coupling matrix in reciprocal space, $K (\bm{k})$.
\begin{equation}
K(\mathbf{k})=\Xi^{-1}(\mathbf{k}).
\end{equation}

Third, the inverse Fourier transformation will be applied to get the elastic coupling matrix in real space,
\begin{equation}
K ^{\mu \Gamma \gamma}_{\mu' \Gamma' \gamma'} (\bm{n})=\frac{1}{N}\sum_{\bm{k}} K ^{\mu \Gamma \gamma}_{\mu' \Gamma' \gamma'} (\bm{k})exp({-i\bm{k} \cdot \bm{n}}).
\end{equation}

Since the contribution of phonon ($K$ and $\Xi$) and uniform strains ($K'$ and $\Xi'$) are independent degrees of freedom, the corresponding total matrix should be the summation of the independent contribution, as shown in Eq. \ref{eq:TotalelasticXi} and Eq. \ref{eq:TotalelasticK}, respectively. 
The procedure of the calculation of $K$ matrix, as shown above, can be applied to either each contribution or the total matrix.

\subsubsection{Lattice Green's functions}
In general, the JT active modes are expressed as:
\begin{equation}
\label{eq:JTactivemode}
q_{\mu\Gamma\gamma}^{\mathbf{n}}=\sum_{\mathbf{n}'\kappa\alpha}c_{\mu\Gamma\gamma}(\mathbf{n}-\mathbf{n}',\kappa,\alpha)\tau_{\kappa\alpha}(\mathbf{n}')
\end{equation}
where expansion coefficients $c_{\mu\Gamma\gamma}(\mathbf{n}-\mathbf{n}',\kappa,\alpha)$ are the displacement of atom $\kappa$ involved in JT active mode $\mu\Gamma\gamma$, which are known for different basic geometries of the centers\cite{Bersuker1989VibronicII}, $\tau_{\kappa\alpha}(\mathbf{n}')$ is the Cartesian displacement of the atom of type $\kappa$ at the unit cell $\mathbf{n}'$ of
the crystal, $\alpha$ denotes the Cartesian axis (x, y, z). The presence of $\mathbf{n}-\mathbf{n}'$ in $c_{\mu\Gamma\gamma}(\mathbf{n}-\mathbf{n}',\kappa,\alpha)$ is due to the fact that there might be atoms from other unit cells involved in one JT center, as the case of the JT center in fcc alkali-doped fullerides A$_3$C$_{60}$ (A is the alkali atom) in which one JT center involves 74 atoms: 60 carbon atoms, 8 cubic alkali atoms, and 6 octahedral alkali atoms, while there are only 63 atoms in the primitive unit cell \cite{Huang2021jahn}.  

In practical, the direct data from the first-principles calculations is the phonon dispersion or dynamic matrices, by which any local distortion can expressed by the phonon modes as\cite{Maradudin1963theory}:
\begin{equation}
\label{eq:phonon2xyz}
\tau_{\kappa\alpha}(\mathbf{n})  =\frac{1}{\sqrt{NM_{\kappa}}}\sum_{\mathbf{k}j}e_{\kappa\alpha}^{\mathbf{k}j}Q_{\mathbf{k}j}exp(i\mathbf{k}\cdot \mathbf{n})
\end{equation}
where $\mathbf{k}$ and $j$ denote the momentum and the branch of the phonon, $M_\kappa$ is the mass of the $\kappa$ atom, $N$ is the number of the unit cells in the crystal, and $e_{\kappa\alpha}^{\mathbf{k}j}$ denotes the polarization vector.
Substituting Eq. \ref{eq:phonon2xyz} into Eq. \ref{eq:JTactivemode}, and comparing it with Eq. \ref{eq:VanVleckexpansion}, we can easily have the expression of the Van Vleck coefficient\cite{Van1940paramagnetic}:
\begin{equation}
\label{eq:expressionVanVleck}
a_{\mu\Gamma\gamma}^{\mathbf{n}}(\mathbf{k}j)=\sum_{\mathbf{n}'\kappa\alpha}\frac{c_{\mu\Gamma\gamma}(\mathbf{n}-\mathbf{n}',\kappa,\alpha)}{\sqrt{NM_{\kappa}}}e_{\kappa\alpha}^{\mathbf{k}j}exp(i\mathbf{k}\cdot \mathbf{n}).
\end{equation}
Substituting Eq. \ref{eq:expressionVanVleck} into Eq. \ref{eq:expre_zeta}, we have the final expression for the $\Xi$ matrix:
\begin{widetext}
\begin{equation}
\label{eq:expr_final_Xi}
\Xi_{\mathbf{n}'\mu'\Gamma'\gamma'}^{\mathbf{n\mu\Gamma\gamma}} =\sum_{\mathbf{l}\mathbf{m}}\sum_{\kappa'\kappa}\sum_{\alpha\beta}c_{\mu'\Gamma'\gamma'}(\mathbf{n}'-\mathbf{l},\kappa',\beta)c_{\mu\Gamma\gamma}(\mathbf{n}-\mathbf{m},\kappa,\beta)G_{\kappa'\alpha,\kappa\beta}(\mathbf{m}-\mathbf{l})
\end{equation}

\end{widetext}
with $G_{\kappa'\alpha,\kappa\beta}(\mathbf{n})$ defined as:
\begin{equation}
\label{expreSLGreenFun}
G_{\kappa'\alpha,\kappa\beta}(\mathbf{n})=\frac{1}{N}\sum_{\mathbf{k}j}\frac{e_{\kappa'\alpha}^{\mathbf{k}j*}e_{\kappa\beta}^{\mathbf{k}j}}{\sqrt{M_{\kappa'}M_{\kappa}}\omega_{\mathbf{k}j}^{2}}exp(i\mathbf{k}\cdot\mathbf{n}).
\end{equation}

This $G_{\kappa'\alpha,\kappa\beta}(\mathbf{n})$ is the LGFs \cite{Flinn1962distortion,Maradudin1963theory}, which can be obtained from both calculations neuron scattering measurements. From  the computational point of view, the LGFs connect the eigenvalues and eigenvectors from the first-principles phonon calculations to the elastic coupling matrix $\Xi$ based on the local JT active modes. On the other hand, the inverse of the LGFs have the same physical meaning of the inter-atomic force constants (IFC) from the standard phonon calculations\cite{Baroni2001phonons}, which is generated from the calculation of the dynamic matrix in the reciprocal space for a grid points in the Brillouin zone (BZ)\cite{Atsushi2023implementation}. However, because of the fact that one JT center might involve more atoms in neighboring unit cells, the evaluation of the elastic coupling for a small grid of unit cells requires the IFC on a larger grid of unit cells. Direct calculation of IFC for larger grid of unit cells is expensive, while a more economic way goes from the IFC on a small grid points in BZ, of which the dynamic matrix is calculated by first-principles method, to generated all the dynamic matrix for a thorough sampling of BZ by Fourier transformation. From this thorough sampling of BZ, the LGFs thus can be computed. 

\subsubsection{Extrema of APES}
The solution of Eq. \ref{eq:Extrema_expre} is the equilibrium values in the extrema points $q_{\mu\Gamma\gamma}^{(\mathbf{n},0)}$, which is expressed as:
\begin{equation}
\label{eq:Q0_expre}
Q_{\overline{\mu}\overline{\Gamma}\overline{\gamma}}^{(0)} =\frac{1}{\omega_{\overline{\mu}\overline{\Gamma}}^{2}}\sum_{\mathbf{n}\mu\Gamma\gamma}a_{\mu\Gamma\gamma}^{\mathbf{n}}(\overline{\mu}\overline{\Gamma}\overline{\gamma})\sum_{\mathbf{n}'\mu'\Gamma'\gamma'}K_{\mathbf{n}'\mu'\Gamma'\gamma'}^{\mathbf{n}\mu\Gamma\gamma}q_{\mu'\Gamma'\gamma'}^{(\mathbf{n}',0)}
\end{equation}
with the stabilization energy as:
\begin{equation}
\label{eq:stabilizationE_0}
U_{\alpha_{1}\alpha_{2}...}  =\sum_{\mathbf{n}\mu\Gamma\gamma}\sum_{\mathbf{n}'\mu'\Gamma'\gamma'}\frac{1}{2}K_{\mathbf{n}'\mu'\Gamma'\gamma'}^{\mathbf{n}\mu\Gamma\gamma}q_{\mu\Gamma\gamma}^{(\mathbf{n},0)}q_{\mu'\Gamma'\gamma'}^{(\mathbf{n}',0)}+\sum_{\mathbf{n}}\epsilon_{\alpha_{\mathbf{n}}}^{\mathbf{n}}(\{q_{\mu\Gamma\gamma}^{\mathbf{n}}\}).
\end{equation}

In general, substituting of $q_{\mu\Gamma\gamma}^{(\mathbf{n},0)}q_{\mu'\Gamma'\gamma'}^{(\mathbf{n}',0)}$ by $q_{\mu\Gamma\gamma}^{\mathbf{n}}q_{\mu'\Gamma'\gamma'}^{\mathbf{n}'}$, the properties of static JT effect around the equilibrium of extrema can be represented, thus the effective Hamiltonian for the static JT effect in multi-center JT systems can be described by:
\begin{equation}
\label{eq:expre_EffectiveHamiltonian}
\hat{H}=\sum_{\mathbf{n}\mu\Gamma\gamma}\sum_{\mathbf{n}'\mu'\Gamma'\gamma'}\frac{1}{2}K_{\mathbf{n}'\mu'\Gamma'\gamma'}^{\mathbf{n}\mu\Gamma\gamma}q_{\mu\Gamma\gamma}^{\mathbf{n}}q_{\mu'\Gamma'\gamma'}^{\mathbf{n}'}+\sum_{\mathbf{n}}U_{\alpha_{\mathbf{n}}}^{\mathbf{n}}(\{q_{\mu\Gamma\gamma}^{\mathbf{n}}\}).
\end{equation}

It should be noted that, the basic idea of such effective Hamiltonian is the downfolding of all vibrational modes or phonons onto the JT active modes, which does not mean that the contribution from any other local distortions other than these JT active modes  is zero. In one word, such effective Hamiltonian can be, in general, used to investigate the APES of any type of the local distortions, such as impurities, defects, etc.  

\subsection{Contribution from uniform strain}
When the concentration of JT centers are low, which means the bulk deformation has nothing to do with JT centers, the above derivation of the effective Hamiltonian is complete.
On the other hand, low-symmetry structural phases of Jahn–Teller crystals can also influence the bulk deformations\cite{Melcher1976anomalous,HZSGehring1975co-operative}. 
Besides, the phonon states and energies should be a function of the strain, however, because of the problem of applying proper boundary conditions when the crystal is stained, it it useful to treat stain separately from the phonons\cite{Kaplan2012cooperative}. 
However, as pointed out by Born and Huang\cite{Born1954dynamical}, the uniform strains describing the bulk deformations of the crystal cannot be reduced to a combination of phonon modes. 
In the following we generalize the approach by taking into account the effects of uniform strains as an independent degree of freedom.

The elastic contribution from uniform strain to the potential energy $U_{us}$ is\cite{Landau1959theory}:
\begin{equation}
\label{eq:Uusgeneral}
U_{us}=\frac{1}{2}NV_{0}\sum_{\alpha\beta,\alpha'\beta'}C_{\alpha\beta,\alpha'\beta'}\varepsilon_{\alpha\beta}\varepsilon_{\alpha'\beta'}
\end{equation}
where N denotes the number of sites, $V_{0}$ is the volume of unit cell, $\varepsilon_{\alpha\beta}$ and $C_{\alpha\beta,\alpha'\beta}$ are general strains and elastic modulus constants with $\alpha,\beta$ representing the Cartesian index ($x$, $y$, $z$). The definition of $\varepsilon_{\alpha\beta}$ is the same as the definition in any text book:
\begin{equation}
\label{eq:starin_definition}
\varepsilon_{\alpha \beta}=\frac{1}{2}\left[\frac{\partial u_\alpha (\mathbf{R})}{\partial R_\beta }+\frac{\partial u_\beta (\mathbf{R})}{\partial R_\alpha }\right],
\end{equation}
where $\mathbf{R}$ is an arbitrary point in the crystal with $\mathbf{u}$ the displacement field. 
For uniform strains, the components $\varepsilon_{\alpha \beta}$ are constants throughout the crystal. 

Same as the treatment of the phonon contribution in Sec. \ref{subsec:phononContribution}, the $U_{us}$ can be expressed by the symmetrized stains and modulus by the transformation with respect to the irreducible representation $\Gamma$ of the corresponding crystalline class as:
\begin{equation}
\label{eq:symmetrizedUniformStrain}
U_{us}=\frac{1}{2}NV_{0}\sum_{\Gamma\gamma}C_{\Gamma\gamma}\varepsilon_{\Gamma\gamma}^{2}
\end{equation}
with the transformation for strain and modulus as:

The uniform strain can be symmetrized by the same transformation as the construction of JT active modes \cite{Landau1959theory}:

\begin{eqnarray}
\label{eq:SymDefStrain}
\varepsilon_{\Gamma\gamma}&=&\sum_{\Gamma\gamma}\Upsilon_{\alpha\beta}^{\Gamma\gamma}\varepsilon_{\alpha\beta} \\
\label{eq:SymDefModuli}
C_{\Gamma \gamma}&=&\sum_{\alpha \beta } \sum_{\alpha' \beta '} C_{\alpha\beta,\alpha'\beta'} \Upsilon_{\alpha \beta} ^{\Gamma \gamma} \Upsilon_{\alpha' \beta'} ^{\Gamma \gamma}
\end{eqnarray}
where $\Upsilon_{\alpha \beta}^{\Gamma \gamma} $ is the unitary transformation from the Cartesian coordinates to the symmetrized basis, and $\varepsilon_{\Gamma\gamma}$, $C_{\Gamma \gamma}$ are the symmetrized strain and the modulus, respectively. 

Based on the Kanamori's work\cite{Kanamori1960crystal}, the uniform strains can be included in the theory of cooperative JT effect as additional terms of vibronic interaction at each JT center. On the other hand, with the symmetrization procedure of general strain and modulus above, the local JT active modes, $\{q_{\overline{\mu}\overline{\Gamma}\overline{\gamma}}^{\mathbf{n}}\}$, can be employed for the full description of the vibronic problems. Thus, the interaction with uniform strains can be included implicitly as additional terms in the Van Vleck expansion, Eq. \ref{eq:expressionVanVleck}, as independent degrees of freedom:
\begin{equation}
\label{eq:VVUStrain}
q_{\mu\Gamma\gamma}^{\mathbf{n}}=\sum_{\mathbf{k}j}a_{\mu\Gamma\gamma}^{\mathbf{n}}(\mathbf{k}j)Q_{\mathbf{k}j} +\sum_{\Gamma'\gamma'}a_{\mu \Gamma\gamma}^{\mathbf{n}}(\Gamma' \gamma')\varepsilon_{\Gamma'\gamma'}.
\end{equation}

Following the procedure as shown in Appendix \ref{sec:appendix_derivation}, the coefficients ($a_{\mu \Gamma\gamma}^{\mathbf{n}}(\Gamma' \gamma')$) of uniform strain contribution has the form:
\begin{equation}
\label{eq:expr_a_UStrain}
a_{\mu \Gamma\gamma}^{us}(\Gamma' \gamma')=\sum_{\alpha\beta} \left[ \sum_{\mathbf{n}'\kappa \alpha } c_{\mu \Gamma\gamma}(\mathbf{n}-\mathbf{n}',\kappa,\alpha)R_{\beta} (\mathbf{n}',\kappa) \right] (\Upsilon^{-1})^{\Gamma'\gamma'}_{\alpha\beta}.
\end{equation}

Since the uniform strain is independent degrees of freedom, Eq. \ref{eq:expre_zeta_main} can be employed directly with the coefficients defined as Eq. \ref{eq:expr_a_UStrain} for the potential contribution from uniform strain:
\begin{equation}
\label{eq:expre_zetaUstrain_main}
\Xi_{\mu'\Gamma'\gamma'}'^{\mu\Gamma\gamma} =\sum_{\overline{\Gamma}\overline{\gamma}}\frac{a_{\mu'\Gamma'\gamma'}(\overline{\Gamma}\overline{\gamma})a_{\mu\Gamma\gamma}(\overline{\Gamma}\overline{\gamma})}{V_0 C^{0}_{\overline{\Gamma}\overline{\gamma}}}.
\end{equation}

Similar to the definition of elastic coupling matrix contributed from phonon, Eq. \ref{eq:KXirelation1}, the elastic coupling matrix contributed from the uniform strains, is calculation as:
\begin{equation}
\label{eq:KdefinitaUStrain}
K'=\Xi'^{-1}.
\end{equation}

\subsection{Total elastic interaction matrix}
After the discussion about the two independent contributions for the elastic interaction for cooperative JT effect, we come to the total expression of the total elastic interaction matrix, expressing as:
\begin{equation}
\label{eq:TotalelasticXi}
\widetilde{\Xi}=\Xi+\frac{1}{N}\Xi',
\end{equation}
with $\Xi$, and $\Xi'$ expressed by Eq. \ref{eq:expr_final_Xi}, and \ref{eq:expre_zetaUstrain_main}, respectively.
The final expression of the elastic interaction for cooperative JT effect is reached as:
\begin{equation}
\label{eq:TotalelasticK}
\widetilde{K}=K+\frac{K'}{N},
\end{equation}
where the phonon contribution ($K$) and the contribution from uniform strains ($K'$) are given by Eq. \ref{eq:KXirelation1}, and Eq. \ref{eq:KdefinitaUStrain} respectively.

\section{Cooperative Jahn-Teller effects in 5d$^1$ double perovskites}
\subsection{Crystal Structure and Jahn-Teller centers}
\begin{figure}
\includegraphics[scale=0.25]{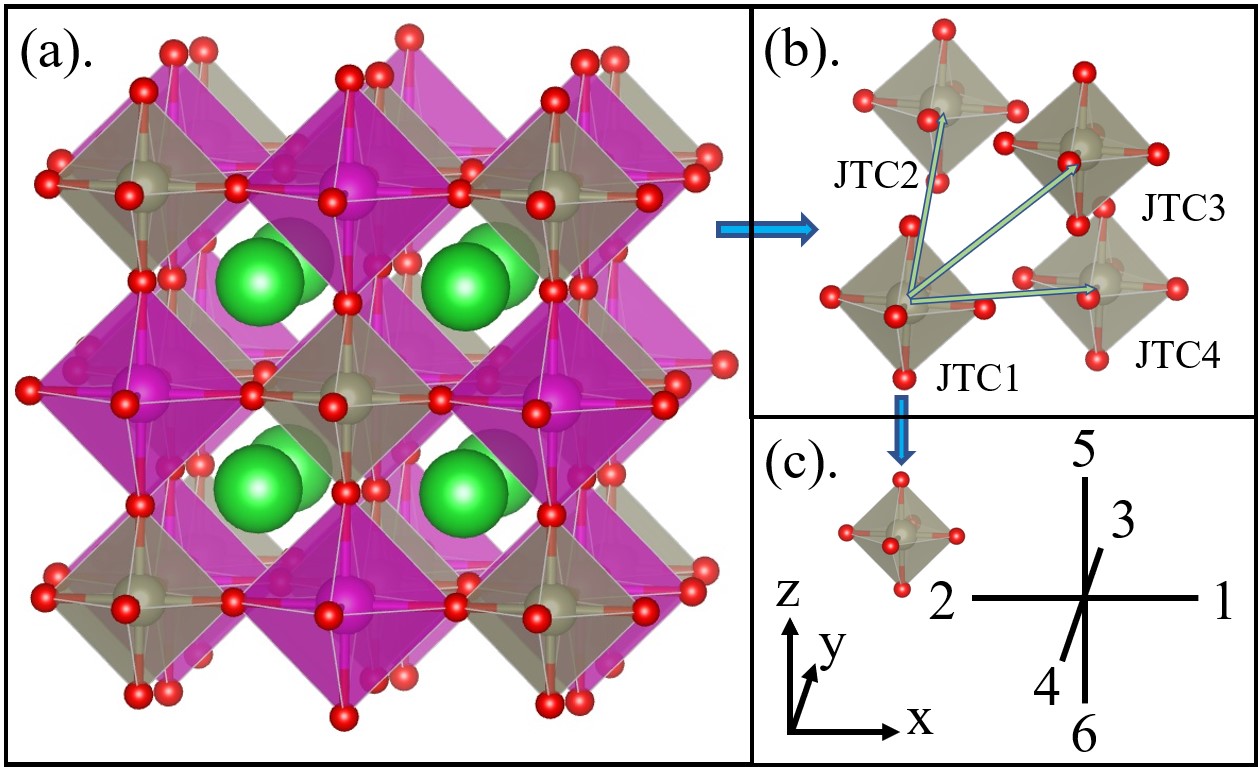}
	\caption{The structure of double perovskites and configuration for Jahn-Teller center. 
	(a). The crystal structure of double perovskite A$_2$B'BX$_6$, green, purple, grey, and red balls are A, B', B and X atoms, respectively. 
	(b). The configuration of four JT centers in convention unit cell. 
	(c). The coordinates and definition of one JT center.
	Detailed information of (b) and (c) can be found in Table. \ref{Table:coordinates_SAJTmodes_Oct}.
	}
\label{Fig:OctahedralConf}
\end{figure}

The interested double perovskites in this paper are: Ba$_2$NaOsO$_6$, Ba$_2$LiOsO$_6$, Ba$_2$MgReO$_6$, Ba$_2$ZnReO$_6$, Cs$_2$TaCl$_6$, Cs$_2$TaBr$_6$ and Ba$_2$YWO$_6$, in which the B site are Ta$^{4+}$, W$^{5+}$, Re$^{6+}$, Os$^{7+}$, with the electron occupation situations 5d$^{1}$ ($T_{2g}^1$).
These double perovskites, A$_2$B'BX$_6$, crystallize into $Fm\overline{3}m$ crystals, with the point group $O_h$, as shown in Fig. \ref{Fig:OctahedralConf} (a), with all the structure parameters shown in Table. \ref{tab:Total_calculation_log_test}, of which the full list of atomic order is shown in Table. \ref{Table:Correspondence_atomicPosition} in the supporting information. The structures are taken from the experiments, either determined by neutron powder diffraction or neutron powder refinements.
However, not all x coefficient of these compounds were provided, such as of Cs$_{2}$TaBr$_{6}$, and Ba$_{2}$YWO$_{6}$, thus x coefficient of these materials were given as a general value, which was 0.2300, not far away from the real situation. A relaxation procedure with force convergence threshold 0.1 eV/$AA$ was further performed to get a consistent x coefficient. 

In this paper, all the calculations were performed for conventional unit cell, including four JT centers (Fig. \ref{Fig:OctahedralConf} (b)). Each JT center consists of the 5$d$ metal atom (Ta, W, Re, and Os) surrounded by six high electronegative atoms (O, Cl or Br), forming an octahedral configuration (\ref{Fig:OctahedralConf} (b)). The label and position of these JT centered atoms are listed in Table.\ref{Table:coordinates_SAJTmodes_Oct}. The relative position of these four JT centers introduces two types of interacting pairs: one is interaction pair of JTC1 and JTC2, and the other one JTC1 and JTC4.
Such an orientation of four JT centers need to be considered, however the analysis will be more complicated, which will be carried out in further studies. Thus in this paper, the relative orientation of the four JT centers will be ignored.

\begin{table}
\caption{\label{tab:Total_calculation_log_test}
The crystal information of the 5$d^1$double perovskites. The space group is $Fm\overline{3}m$, with the Wyckoff position for atomic positions are A: 8c(1/4, 1/4, 1/4), B': 4b(1/2,0,0), B: 4a(0, 0, 0), and X: 24e(x,0,0),
}
\begin{tabular}{|c|c|c|c|c|c|c|c|}
\hline
\multirow{2}{*}{A} &\multirow{2}{*}{B'} &\multirow{2}{*}{B} & \multirow{2}{*}{X} &\multirow{2}{*}{A$_2$(B')BX$_6$} & \multicolumn{2}{c|}{Structure} &  \multirow{2}{*}{Config}\\
\cline{6-7}
&  & & &              &  a/$\AA$ & x(O)   &  \\
\hline
\hline
\multirow{5}{*}{Ba}&Na&\multirow{2}{*}{Os} & \multirow{5}{*}{O} & Ba$_{2}$NaOsO$_{6}$\cite{Stitzer2002Crystal}  & 8.2870  & 0.2256 &    \multirow{7}{*}{5d$^{1}$} \\
\cline{2-2} \cline{5-7}
& Li & & &Ba$_{2}$LiOsO$_{6}$\cite{Stitzer2002Crystal} & 8.1046  & 0.2330 &     \\
\cline{2-3} \cline{5-7}
& Mg & \multirow{2}{*}{Re}& &Ba$_{2}$MgReO$_{6}$\cite{Marjerrison2016Cubic}  & 8.0849 & 0.2392 &\\
\cline{2-1} \cline{5-7}
& Zn & & &Ba$_{2}$ZnReO$_{6}$\cite{daCruz2022Impact}     & 8.1061 & 0.23952 &  \\
\cline{2-3} \cline{5-7}
& Y &W & &Ba$_{2}$YWO$_{6}$\cite{Lee2021Experimental} & 8.3848 & 0.2350 & \\
\cline{1-7}
\multirow{2}{*}{Cs}& &\multirow{2}{*}{Ta} &Cl   & Cs$_{2}$TaCl$_{6}$\cite{Yun2006Dicaesium}  & 10.2710 & 0.2331 &  \\
\cline{4-7}
& & & Br & Cs$_{2}$TaBr$_{6}$\cite{Ishikawa2021Ligand} & 10.7685 & 0.2394 & \\
\hline
\end{tabular}
\end{table}

Since there is only one electron, due to the crystal field, the 5$d$ orbital splits into $E_{g}$ and $T_{2g}$, with $E_{g}$ higher than $T_{2g}$. The JT involved orbitals are $T_{2g}$. The existence of spin-orbital interaction splits $T_{2g}$ into $\Gamma_7$ and $\Gamma_8$. The low lying $\Gamma_8$ is the target orbital. The irreducible representation for this system is given as $\Gamma_{vib}^{oct}=a_{1g}\bigoplus e_{g}\bigoplus t_{2g} \bigoplus t_{1u}\bigoplus t_{2u}$, therefore according to selection rule,
the $\Gamma_{8}$ orbitals linearly couple to JT active modes involved in the symmetric product of $\Gamma_{8}$ representation: $[\Gamma_{8}^2]=a_{1g}\bigoplus e_{g}\bigoplus t_{2g}$. $a_{1g}$ will not be considered because of the fact that it only shifts the energy levels without any orbital splitting, thus all the parameters corresponding to $a_{1g}$ will be ignored in the following analysis. 

\begin{table}
\caption{\label{Table:coordinates_SAJTmodes_Oct}
Coordinates and symmetry adapted JT active modes, $q_{\mu \Gamma \gamma}$, of octahedral configuration. $\mu$ represents the position of JT center, labeled by JTC1 to JTC4 for the four JT centers in the convention unit cell, while $\Gamma\gamma$ distinguish JT active modes with the label JTD\_1 to JTD\_6, corresponding to $a_{1g}$, $e_{g\theta}$, $e_{g\epsilon}$, $t_{2g\xi}$, $t_{2g\eta}$, and $t_{2g\zeta}$.}
\resizebox{0.5\textwidth}{!}{
\begin{tabular}{|c|c|c|c|c|c|c|c|}
\hline
\multicolumn{8}{|c|}{JT active mode $q_{\mu \Gamma \gamma}$ for octahedral configuration} \\
\hline
Atoms & Position & $a_{1g}$ & \multicolumn{2}{c|}{$e_{g}$} & \multicolumn{3}{c|}{$t_{2g}$} \\
\hline
&  &  & $\theta$ & $\epsilon$ & $\xi$ & $\eta$ & $\zeta$ \\
\cline{3-8}
& & JTD\_1 & JTD\_2 & JTD\_3 & JTD\_4 & JTD\_5 & JTD\_6 \\
\hline
A1 & (1,0,0) & $\frac{1}{\sqrt{6}}(1,0,0)$ & $\frac{1}{2\sqrt{3}}(-1,0,0)$ & $\frac{1}{2}(1,0,0)$ & $(0,0,0)$ & $\frac{1}{2}(0,0,1)$ & $\frac{1}{2}(0,1,0)$ \\
\hline
A2 & (-1,0,0) & $\frac{1}{\sqrt{6}}(-1,0,0)$ & $\frac{1}{2\sqrt{3}}(1,0,0)$ & $\frac{1}{2}(-1,0,0)$ & $(0,0,0)$& $\frac{1}{2}(0,0,-1)$ & $\frac{1}{2}(0,-1,0)$ \\
\hline
A3 & (0,1,0) & $\frac{1}{\sqrt{6}}(0,1,0)$ & $\frac{1}{2\sqrt{3}}(0,-1,0)$ & $\frac{1}{2}(0,-1,0)$ &  $\frac{1}{2}(0,0,1)$ & $(0,0,0)$ & $\frac{1}{2}(1,0,0)$ \\
\hline
A4 & (0,-1,0) & $\frac{1}{\sqrt{6}}(0,-1,0)$ & $\frac{1}{2\sqrt{3}}(0,1,0)$ & $\frac{1}{2}(0,1,0)$ &  $\frac{1}{2}(0,0,-1)$ & $(0,0,0)$ & $\frac{1}{2}(-1,0,0)$ \\
\hline
A5 & (0,0,1) & $\frac{1}{\sqrt{6}}(0,0,1)$ & $\frac{1}{2\sqrt{3}}(0,0,2)$ & $(0,0,0)$ &  $\frac{1}{2}(0,1,0)$ &   $\frac{1}{2}(1,0,0)$ & $(0,0,0)$ \\
\hline
A6 & (0,0,-1) & $\frac{1}{\sqrt{6}}(0,0,-1)$ & $\frac{1}{2\sqrt{3}}(0,0,-2)$ & $(0,0,0)$ & $\frac{1}{2}(0,-1,0)$  & $\frac{1}{2}(-1,0,0)$  & $(0,0,0)$ \\
\hline
\hline
\multicolumn{8}{|c|}{JT active center}\\
\hline
JTC1 & \multicolumn{7}{c|}{$(0,0,0)$}\\
\hline
JTC2 & \multicolumn{7}{c|}{$(0,0.5,0.5)$}\\
\hline
JTC3 & \multicolumn{7}{c|}{$(0.5,0,0.5)$}\\
\hline
JTC4 & \multicolumn{7}{c|}{$(0.5,0.5,0)$}\\
\hline
\end{tabular}
	}
\end{table}

\subsection{Computational methods}
\label{subsec:Computational}
All the DFT calculations were performed by VASP\cite{VASP_1,VASP_2,VASP_3,VASP_4}. The plane-wave kinetic energy cutoff was set to 400 eV for Ba$_2$NaOsO$_6$, Ba$_2$LiOsO$_6$, Ba$_2$MgReO$_6$, Ba$_2$ZnReO$_6$, and Ba$_2$YWO$_6$, with 300 eV for Cs$_2$TaCl$_6$, and Cs$_2$TaBr$_6$, and a $6 \times 6 \times 6$ Monkhorst-Pack mesh was used to perform Brillouin zone (BZ) integration in order to ensure the convergence of self-consistent field calculation.
The convergence threshold of total energy was set to 10$^{-7}$ eV with that of the forces on atoms to 10$^{-4}$ eV/$\AA$ for full relaxation calculations. 
Projector augmented-wave(PAW) pseudo-potentials \cite{PAW_first,PAW_VASP} are employed with the exchange and correlation approximations Generalized Gradient Approximation PBE (GGA-PBE) \cite{PBE-1,PBE-2}.

The phonon calculations were carried out by frozen-phonon method with VASP+phonopy\cite{Togo2023first_phonopy,Togo2023implementation_phonopy}, in which a 2 $\times$ 2 $\times$ 2 super-cell was used, with the energy convergence threshold set as 10$^{-9}$ eV. The LGFs are generated on a $31 \times 31 \times 31$, Monkhorst-Pack meshes in BZ. The elastic modulus was calculated by the density functional perturbation theory (DFPT)\cite{Le2002symmetry, Wu2005systematic}, with a $12 \times 12 \times 12$ Monkhorst-Pack mesh to guarantee the convergence.
\subsection{$\Xi$ matrix}
\subsubsection{Phonon dispersion}
\begin{figure}
\includegraphics[scale=0.18]{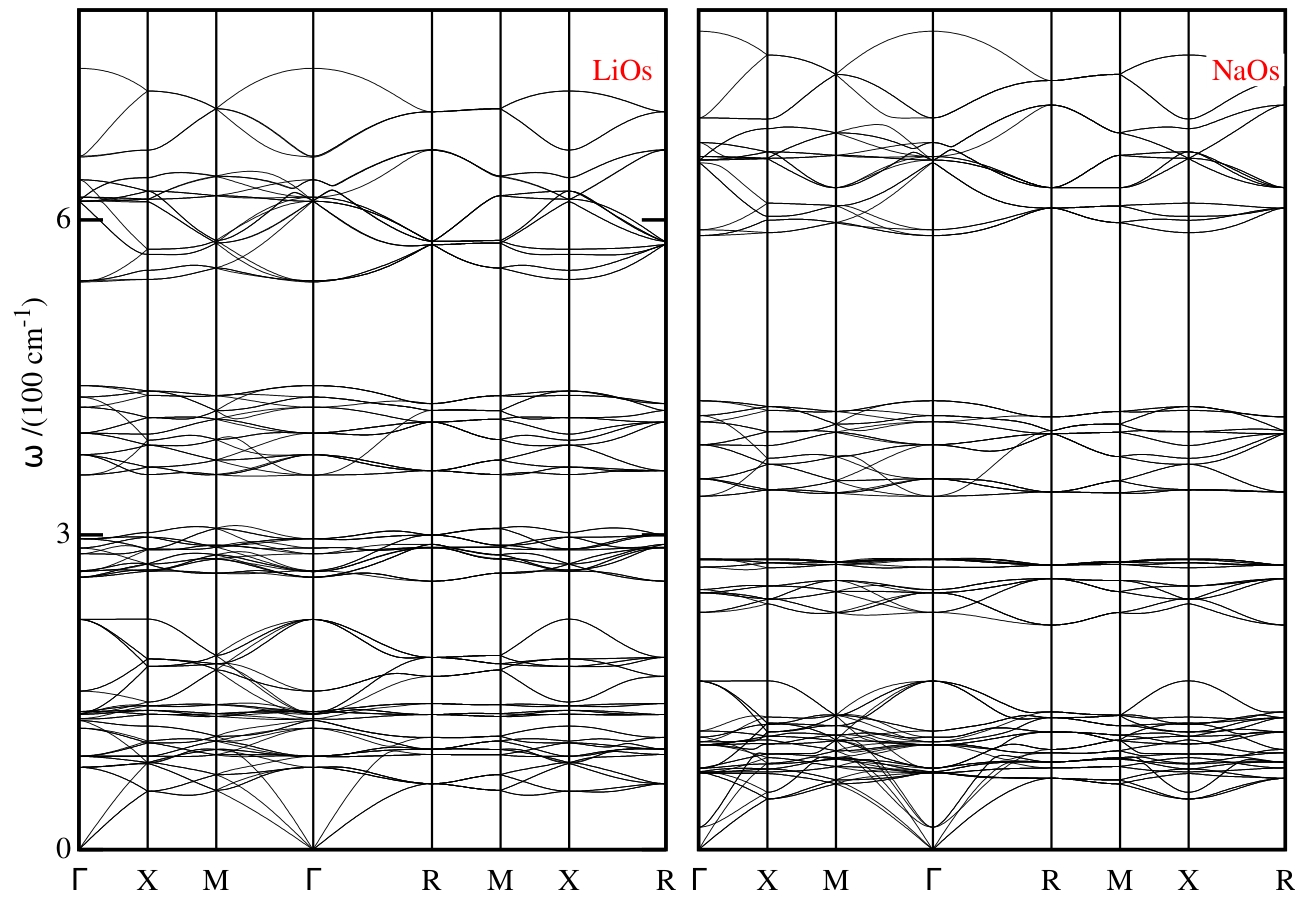}
\caption{Phonon dispersions of for Ba$_2$LiOsO$_6$ and Ba$_2$NaOsO$_6$, distinguished by `LiOs' and `NaOs', respectively.}
\label{Fig:phonondispersion}
\end{figure}

The phonon dispersions of Ba$_2$LiOsO$_6$ and Ba$_2$NaOsO$_6$ are shown in Fig. \ref{Fig:phonondispersion}, while the phonon dispersions of all the 7 double perovskites can be found in Fig.\ref{Fig:phonondispersion_all} in supporting information. Ba$_2$NaOsO$_6$ has more broadening dispersion than Ba$_2$LiOsO$_6$ because of the large ratio between the cations and anions. This is clearly from the phonon density of states (DOS), where O atoms mainly contribute to the high energy phonon DOS. 

\subsubsection{$\Xi$ matrix from phonons}
The key parameters in phonon calculation are the IFCs, which describes the response property of one atom with respect to another atom in the crystal. The real space dimension of IFC corresponds to the super-cell generated in the phonon calculation procedure. From the IFCs, by Fourier transformation, the LGFs are generated. The $\Xi$ matrix are computed by the LGFs and the JT active modes, defined in Table. \ref{Table:coordinates_SAJTmodes_Oct}, by Eq. \ref{eq:expr_final_Xi}. 

The calculated $\Xi$ matrix is shown in Table. \ref{tab:ECCs_summary_all_includingUS}. 
The onsite coupling parameters are diagonal and have a two and three degeneracy, corresponding to the $E_g$ and $T_{2g}$ states. 
\input{./Eccs_summary_all_US }

\subsubsection{$\Xi$ matrix from uniform strain}
The evaluation of $\Xi$ matrix from uniform strain are carried out with Eq. \ref{eq:expre_zetaUstrain_main}.
The required parameters, such as volumes of the convention unit cell and the elastic modulus are listed in Table. \ref{Table:ModuliC_all}.
The $\Xi$ matrix for Ba$_2$NaOsO$_6$ and Ba$_2$LiOsO$_6$ from the uniform strain, are shown in Table. \ref{tab:ECCs_summary_all_includingUS}. 
Similarly to that of phonon, the $\Xi$ matrix of uniform strain for the onsite JT center is diagonal with two and three degeneracy. Besides, they have the same magnitude order. However, the LGFs are generally expected to decrease fast with the distance between the centers, as $(\mathbf{n}-\mathbf{n}')^{-3}$\cite{Orbach1967phonon}, so as the $\Xi$ matrix. While the contribution from uniform strains remains constant for any distance between JT centers. Therefore phonons and uniform strains contribute comparably to the interaction energy of a given JT center with all other centers of the crystal. From this aspect, it is important to include the influence of uniform strain for Cooperative JT effect. 

\begin{table}
\caption{\label{Table:ModuliC_all}
The volume $V_0$ (1000 $\AA^3$) and three independent general modulus, $C_{xxxx}$, $C_{xxyy}$, and $C_{xyxy}$ (10$^{3}$ kBar) as well as the corresponding symmetrized three independent modulus, $C_{A_1}$, $C_{E_{\theta}}$, and $C_{T_{2(\xi)}}$. `LiOs', `NaOs', `MgRe', `ZnRe', `YW', `TaCl', and `TaBr' represent for Ba$_2$NaOsO$_6$, Ba$_2$LiOsO$_6$, Ba$_2$MgReO$_6$, Ba$_2$ZnReO$_6$, Cs$_2$TaCl$_6$, Cs$_2$TaBr$_6$,  and Ba$_2$YWO$_6$.
}
\begin{tabular}{|c|c|ccc|ccc|}
\hline
	& $V_0$ & $C_{xxxx}$ & $C_{xxyy}$ & $C_{xyxy}$ & $C_{A_1}$ & $C_{E_{\theta}}$ & $C_{T_{2(\xi)}}$ \\
\hline
 NaOs & 0.5691 & 2.291 & 0.727 & 0.668 & 3.746 & 1.564 & 2.672 \\
 \hline
 LiOs & 0.5323 & 2.027 & 0.921 & 0.878 & 3.868 & 1.106 & 3.512 \\
 \hline
 MgRe & 0.5285 & 2.889 & 1.115 & 1.133 & 5.119 & 1.774 & 4.532 \\
 \hline
 ZnRe & 0.5326 & 2.678 & 1.340 & 1.022 & 5.358 & 1.338 & 4.088 \\
 \hline
 TaCl & 1.0835 & 0.264 & 0.126 & 0.106 & 0.516 & 0.139 & 0.426 \\
 \hline
 TaBr & 1.2487 & 0.203 & 0.101 & 0.088 & 0.405 & 0.101 & 0.352 \\
 \hline
 YW   & 0.5895 & 3.603 & 0.943 & 0.811 & 5.489 & 2.661 & 3.245 \\
 \hline
   \end{tabular}
\end{table}

\subsection{Effective elastic coupling matrix $K$}
Based on the method described in Sec. \ref{sec:appendix_effective}, the effective elastic coupling matrix, $K$, can be obtain for both phonons and uniform strains. 

Though there are two types JT active modes ($E_g$, and $T_{2g}$) on each JT center, only $E_g$ modes are considered, as the vibronic coupling constants of $E_g$ are 6 times larger than that of $T_{2g}$ according to N. Iwahara's study\cite{Iwahara2018spin},
meaning that the contribution of $E_g$ to JT energy is 36 larger than that of $T_{2g}$. Thus in the analysis of elastic coupling constants, $T_{2g}$ JT active modes are ignored for simplicity.

In the previous section, $\Xi ^{\mu \gamma}_{\mu' \gamma'} (\bm{R})$ ($\mu,\mu'$ donate the four JT centers, $\gamma, \gamma' $ donate the two components, $\theta, \epsilon$ of $E_g$ JT active mode, and $\bm{R}$ represents the lattice vector) has been truncated to the 4th next nearest conventional unit cells, due to the fast decay of $\Xi ^{\mu \gamma}_{\mu' \gamma'} (\bm{R})$ on $\bm{R}$. Thus, the elastic coupling matrix $K$ can also be obtained for up to the 4th next nearest conventional unit cells. 

The $K$ matrix of $E_g$ modes for Ba$_2$LiOsO$_6$ and Ba$_2$NaOsO$_6$ from the all-phonon contribution is summarized in 
Table. \ref{tab:K_real_summary_EgB11}. These parameters can also be obtained based on the IFC from first-principles phonon calculations. However, as demonstrated in the Sec. \ref{subsec:Computational}, the $2\times 2\times 2$ q-mesh is used for the phonon calculations, meaning that only IFCs on a grid of $2\times 2\times 2$ can be generated. While the JT centers usually involves in atoms in more than one unit cell, the direct calculation of $K$ matrix from IFC limits the analysis of the distance-dependence of elastic coupling. 
\input{./B11Kappa_real_summary_Latex_8_Eg}
A full list of $K_{g}$ matrix of $E_g$ modes for different unit cells 
is shown in Table. \ref{tab:K_real_summary_allB11}
in the supporting information.
On the other hand, for the contribution of uniform strain, \ref{tab:K_real_summary_allUS2}.
\input{./US2Kappa_real_summary_Latex_8}

\section{Discussion}
\begin{figure}
	\includegraphics[scale=0.31]{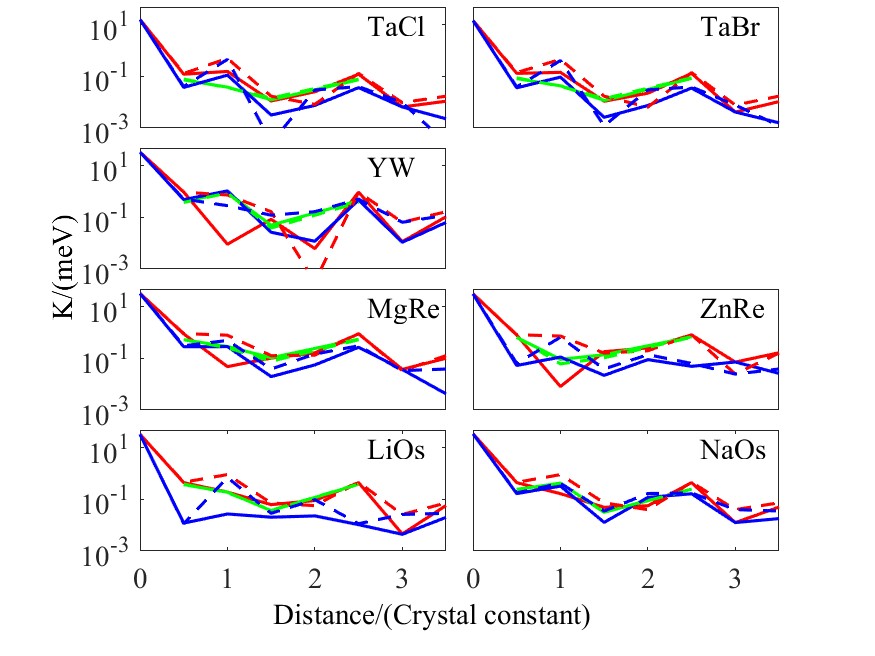}
	\caption{Distance dependence of K based on all phonons. Red, green and blue represent the parameter $K^{E(\theta)}_{E(\theta)}(\mathbf{n} )$, $K^{E(\theta)}_{E(\epsilon)}(\mathbf{n})$ and $K^{E(\epsilon)}_{E(\epsilon)}(\mathbf{n})$, respectively. The solid and dashed lines represents the process of calculating K matrix with and without the influence of periodicity. 
	}
	\label{Fig:K_distance_phonon}
\end{figure}

In order to have a straightforward comparison with other interaction strength, the elastic coupling matrix $K$ is converted with respect to dimensionless JT active modes via $K/ \sqrt{(\omega_{\Gamma\gamma}\omega_{\Gamma'\gamma'}M^{am}_{\Gamma\gamma}M^{am}_{\Gamma'\gamma'}}$,), where $M^{am}$ and $\omega_{\Gamma\gamma}$ are the mass of the corresponding atoms in JT mode $\Gamma\gamma$, and the vibration frequencies of $\Gamma\gamma$ JT active modes calculated by the $K$ matrix, as shown in Table. \ref{Table:Omega_K}.
\input{./B11Summary_freq_Eg}
As discussed above, the $\Xi$ matrix has $(\mathbf{n}-\mathbf{n}')^{-3}$ decay with respect to the distance between JT centers, which should be the same situation for elastic matrix $K$. The distance dependence of elastic coupling parameters ($K^{\mathbf{n}\Gamma\gamma}_{\mathbf{n}'\Gamma'\gamma'}$) in Fig. \ref{Fig:K_distance_phonon} (solid line). This is the effective $K$ matrix elements including the influence of surrounding JT centers in the crystals. On the other hand, in the extreme situation where there is only one JT impurity or there are only two JT impurities. The interactions between the two JT impurities and the JT active modes on the only one JT impurity become simpler, of which the procedure is shown in \ref{sec:appendix_IsolatedJTimp}, different from that described in \ref{sec:appendix_effective}. The results of such JT impurities with respect to distances are shown in Fig. \ref{Fig:K_distance_phonon} (dashed line). 

For all the elastic matrix, from phonons or uniform strain, the onsite parameters have the degeneracy of 2, corresponding to the $E_g$ JT active modes. To evaluate the dynamic situation, we calculate the contribution of the inter site JT energy and compared to the previous study\cite{Iwahara2018spin}, which is carried out with quantum chemistry computations, as shown in Table. \ref{tab:EJT_intersite}. $E_{JT}^{E}$ is the amplitude of the JT stabilization of inter site case on the trough is about 80 times larger than that of inter site case, which indicates unquenched dynamic JTE onsite in these systems.

The $E_g$ JT modes frequencies of Ba$_2$MgWO$_6$ are about 540 $cm^{-1}$ based on the Raman spectra \cite{HZSPasztorova2023structure}, which should have similar value as Ba$_2$MgWO$_6$. Our prediction for $E_g$ JT modes frequencies of Ba$_2$MgWO$_6$ is 543.886 $cm^{-1}$, which appears close to that from experiment. 

\begin{table}
\caption{\label{tab:EJT_intersite}
The elastic energy between nearest JT centers. 
}
\begin{tabular}{|c|c|c|}
\hline
&Ba$_{2}$NaOsO$_{6}$ & Ba$_{2}$LiOsO$_{6}$ \\
\hline
$q_{E\gamma}^{\mathbf{n} (\mathbf{0})}$ (a.u.)    &    $10.24$   &    $12.29$    \\
\hline
$E_{JT}^{E}$ (meV)      &    $20.31$        &    $32.22$         \\
\hline
$K^{\mathbf{n}}_{\mathbf{n}'}(q_{E\gamma}^{\mathbf{n} (\mathbf{0})})^2$ (meV) : $ \left( \begin{array}{cc}
    \theta \theta   & \theta \epsilon  \\
    \epsilon \theta & \epsilon \epsilon  \\ 
\end{array}\right) $
& $ \left( \begin{array}{cc}
    0.52  &  0.28 \\
    0.28  &  0.19 \\ 
\end{array}\right) $
& $ \left( \begin{array}{cccc}
    0.85  &  0.72  \\
    0.72  &  0.02  \\  
\end{array}\right) $ \\
\hline
\end{tabular}
\end{table}

\section{CONCLUSIONS}
In this paper, an effective Hamiltonian for a static cooperative JT effect is proposed, of which the all the necessary microscopic parameters can be obtained from first-principles calculations. This methodology is applied to a series of double perovskites A$_2$B'BX$_6$. 
This methodology can be, in general, used to investigate the APES of any type of the local distortions, such as impurities, defects, etc
\section*{Acknowledgements}
Z.H. acknowledges the financial support of the research projects A-800079-00-00 and A-8000017-00-00 of the National University of Singapore. 
N.I. acknowledges the Grant-in-Aid for Scientific Research (Grant No. 22K03507) from the Japan Society for the Promotion of Science.
The computational resources and services used in this work were provided by the VSC (Flemish Supercomputer Center) funded by the Research Foundation - Flanders (FWO) and the Flemish Government - department EWI.

\appendix
\section{Derivation of the effective elastic interaction matrix}
\label{sec:appendix_effectiveElastic}
Considering the following expression based on Eq. \ref{eq:VanVleckexpansion}, and Eq. \ref{eq:APES_total}, 

\begin{eqnarray}
\label{eq:LagrangeMF}
F_{\alpha_{1}\alpha_{2}...}=&&U_{\alpha_{1}\alpha_{2}...} +  \nonumber \\  &&\sum_{\mathbf{n}}\sum_{\overline{\mu}\overline{\Gamma}\overline{\gamma}}\lambda_{\overline{\mu}\overline{\Gamma}\overline{\gamma}}^{\mathbf{n}}\left(q_{\overline{\mu}\overline{\Gamma}\overline{\gamma}}^{\mathbf{n}}-\sum_{\mu\Gamma\gamma}a_{\overline{\mu}\overline{\Gamma}\overline{\gamma}}^{\mathbf{n}}(\mu\Gamma\gamma)Q_{\mu\Gamma\gamma}\right)
\end{eqnarray}
with $\lambda_{\mu\Gamma\gamma}^{(\mathbf{n})}$ the Lagrange multipliers, the extrema equations are:

\begin{eqnarray}
\label{eq:LagrangeMF_1}
\frac{\partial F_{\alpha_{1}\alpha_{2}...}}{\partial Q_{\mu\Gamma\gamma}} & =&\omega_{\mu\Gamma}Q_{\mu\Gamma\gamma}-\sum_{\mathbf{n}}\sum_{\overline{\mu}\overline{\Gamma}\overline{\gamma}}\lambda_{\overline{\mu}\overline{\Gamma}\overline{\gamma}}^{\mathbf{n}}a_{\overline{\mu}\overline{\Gamma}\overline{\gamma}}^{\mathbf{n}}(\mu\Gamma\gamma)=0\\
\label{eq:LagrangeMF_2}
\frac{\partial F_{\alpha_{1}\alpha_{2}...}}{\partial q_{\overline{\mu}\overline{\Gamma}\overline{\gamma}}^{\mathbf{n}}} & =&\frac{\partial\epsilon_{\alpha}^{\mathbf{n}}(\{q_{\overline{\mu}\overline{\Gamma}\overline{\gamma}}^{\mathbf{n}}\})}{\partial q_{\mu{\Gamma}\overline{\gamma}}^{(\mathbf{n})}}+\lambda_{\overline{\mu}\overline{\Gamma}\overline{\gamma}}^{\mathbf{n}}=0\\
\label{eq:LagrangeMF_3}
\frac{\partial F_{\alpha_{1}\alpha_{2}...}}{\partial\lambda_{\overline{\mu}\overline{\Gamma}\overline{\gamma}}^{\mathbf{n}}} & =&q_{\overline{\mu}\overline{\Gamma}\overline{\gamma}}^{\mathbf{n}}-\sum_{\mu\Gamma\gamma}a_{\overline{\mu}\overline{\Gamma}\overline{\gamma}}^{\mathbf{n}}(\mu\Gamma\gamma)Q_{\mu\Gamma\gamma}=0.
\end{eqnarray}
With the Eq. \ref{eq:LagrangeMF_3}, and multiplying $a_{\overline{\mu}\overline{\Gamma}\overline{\gamma}}^{\mathbf{n}}(\mu\Gamma\gamma)/\omega^2_{\Gamma\gamma}$ to Eq. \ref{eq:LagrangeMF_1}, we have:
\begin{equation}
q_{\overline{\mu}\overline{\Gamma}\overline{\gamma}}^{\mathbf{n}}  =  \sum_{\mathbf{n}'\mu'\Gamma'\gamma'}\lambda_{\mu'\Gamma'\gamma'}^{\mathbf{n}'}\Xi_{\mathbf{n}'\mathop{\mu'\Gamma'\gamma'}}^{\mathbf{n\overline{\mu}\overline{\Gamma}\overline{\gamma}}}
\end{equation}
where $\Xi_{\mathbf{n}'\mu'\Gamma'\gamma'}^{\mathbf{n\overline{\mu}\overline{\Gamma}\overline{\gamma}}}$ is defined as:
\begin{equation}
\label{eq:expre_zeta}
\Xi_{\mathbf{n}'\mu'\Gamma'\gamma'}^{\mathbf{n\overline{\mu}\overline{\Gamma}\overline{\gamma}}}  =  \sum_{\mathbf{k}j}\frac{a_{\mu'\Gamma'\gamma'}^{\mathbf{n}'}(\mathbf{k}j)a_{\overline{\mu}\overline{\Gamma}\overline{\gamma}}^{\mathbf{n}}(\mathbf{k}j)}{\omega^2_{\mathbf{k}j}}.
\end{equation}

Since the matrix $\Xi$ consisting of the elastic parameters of the multi-atomic system only, is Hermitian, of which the inverse matrix ($K=\Xi^{-1}$) is Hermitian as well, the Lagrange multipliers can be expressed as 
\begin{equation}
\label{eq:lambda}
\lambda_{\overline{\mu}\overline{\Gamma}\overline{\gamma}}^{\mathbf{n}} =\sum_{\mathbf{n}'\mu'\Gamma'\gamma'}K_{\mathbf{n}'\mu'\Gamma'\gamma'}^{\mathbf{n\overline{\mu}\overline{\Gamma}\overline{\gamma}}}q_{\mu'\Gamma'\gamma'}^{\mathbf{n}'}.
\end{equation}
The extrema expression based on only the local JT active modes can be obtained by substituting the expression Eq. \ref{eq:lambda} in to Eq. \ref{eq:LagrangeMF_2}, which goes like:
\begin{equation}
\label{eq:Extrema_expre_appendix}
\frac{\partial\epsilon_{\alpha_{\mathbf{n}}}^{\mathbf{n}}(\{q_{\overline{\mu}\overline{\Gamma}\overline{\gamma}}^{\mathbf{n}}\})}{\partial q_{\overline{\mu}\overline{\Gamma}\overline{\gamma}}^{\mathbf{n}}}+\sum_{\mathbf{n}'\mu'\Gamma'\gamma'}K_{\mathbf{n}'\mu'\Gamma'\gamma'}^{\mathbf{n\overline{\mu}\overline{\Gamma}\overline{\gamma}}}q_{\mu'\Gamma'\gamma'}^{\mathbf{n}'}  =  0
\end{equation}

\section{Derivation of coefficients $a_{\mu\Gamma\gamma}^{\mathbf{n}}(\Gamma \gamma)$ }
\label{sec:appendix_derivation}
The atomic displacements $u_{\kappa\alpha}(\mathbf{n})$can be decomposed into the contributions from phonons ($u^{ph}_{\kappa\alpha}(\mathbf{n})$) and uniform strains ($u^{us}_{\kappa\alpha}(\mathbf{n})$), as 
\begin{equation}
\label{eq:DisplacementDecom}
u_{\kappa\alpha}(\mathbf{n})=u^{ph}_{\kappa\alpha}(\mathbf{n})+u^{us}_{\kappa\alpha}(\mathbf{n}).
\end{equation}

The uniform strain terms can be expressed as:
\begin{equation}
\label{eq:UStrainDisplace}
u^{us}_{\kappa\alpha}(\mathbf{n})=\sum_{\beta}\widetilde{\varepsilon}_{\alpha \beta} R_{\alpha} (\mathbf{n}\kappa)
\end{equation}
where $\mathbf{R}(\mathbf{n}\kappa)$ is the position of the atom ($\mathbf{n}\kappa$) in a common coordinate system and the coefficients $\widetilde{\varepsilon}_{\alpha \beta}=\partial u^{us}_{\alpha}/ \partial R_{\beta}$ are constants in the case of uniform strains, of which the total number is nine. The nine coefficients includes 6 symmetric tensor elements, which are equivalent to the six tensor elements of the general strain, and 3 anti-symmetric tensor elements, describing the uniform bulk rotation of the crystal around three Cartesian axes. Thus, for the cooperative JT problem, the general strain tensor can be used as the coefficients $\widetilde{\varepsilon}_{\alpha \beta}$. Thus, Eq. \ref{eq:UStrainDisplace}, can be written as:
\begin{equation}
\label{eq:UStrainDisGeneralTensor}
u^{us}_{\kappa\alpha}(\mathbf{n})=\sum_{\beta}\varepsilon_{\alpha \beta} R_{\alpha} (\mathbf{n}\kappa).
\end{equation}

With the definition of JT active modes, Eq. \ref{eq:JTactivemode}, the uniform strains contribution, the second term in Eq. \ref{eq:VVUStrain}, becomes
\begin{equation}
\label{eq:ContriUStrain2JTmodes}
q'^{\mathbf{n}}_{\mu\Gamma\gamma}=\sum_{\alpha\beta} \left[ \sum_{\mathbf{n}'\kappa \alpha } c_{\mu \Gamma\gamma}(\mathbf{n}-\mathbf{n}',\kappa,\alpha)R_{\beta} (\mathbf{n}',\kappa) \right] \varepsilon_{\alpha\beta}.
\end{equation}
The translational symmetry of the uniform strains results in the invariance of the coefficients $c_{\mu \Gamma\gamma}(\mathbf{n}-\mathbf{n}',\kappa,\alpha)$, and $\sum_{\mathbf{n}'\kappa \alpha } c_{\mu \Gamma\gamma}(\mathbf{n}-\mathbf{n}',\kappa,\alpha)=0$, which are valid for all the JT active modes. With the transformation between general strain and the symmetric strain, Eq. \ref{eq:SymDefStrain}, we can obtain the expression of the coefficients for the uniform strain contribution, Eq. \ref{eq:expr_a_UStrain}, in which the $\mathbf{n}$ is omitted because of the translational invariance.

\section{Derivation of the symmetrized elastic modulus for cubic systems}
\label{sec:appendix_symmetrized}
Starting from Eq. \ref{eq:Uusgeneral}, for cubic system, there are only three independent elastic modulus element, of which the relations of all the element of elastic modulus is shown as following:
\begin{eqnarray}
C_{xxxx} & =&C_{yyyy}=C_{zzzz}\nonumber \\
C_{xxyy} & =&C_{xxzz}=C_{yyzz}=C_{zzyy}=C_{zzxx}=C_{yyxx}\nonumber \\
C_{xyxy} & =&C_{yzyz}=C_{zxzx}\nonumber \\
C_{xyzx} & =&C_{xyxx}=C_{xyyy}=C_{xyzz}=C_{xyyz}=0\nonumber \\
C_{yzxx} & =&C_{yzyy}=C_{yzzz}=C_{xzxy}=C_{yzzx}=0\\
C_{zxxx} & =&C_{zxyy}=C_{zxzz}=C_{zxxy}=C_{zxyz}=0\nonumber \\
C_{xxxy} & =&C_{xxyz}=C_{xxzx}=C_{yyxy}=C_{yyyz}\nonumber \\
         & =&C_{yyzx}=C_{zzxy}=C_{zzyz}=C_{zzzx}=0\nonumber 
\end{eqnarray}
while $\varepsilon_{xy}=\varepsilon_{yx}$, $\varepsilon_{yz}=\varepsilon_{zy}$, and $\varepsilon_{zx}=\varepsilon_{xz}$, so that the Eq. \ref{eq:Uusgeneral} can be expanded as 

\begin{eqnarray}
\label{eq:UusLandau}
U_{us}&=&\frac{NV_{0}}{2}C_{xxxx}(\varepsilon_{xx}^{2}+\varepsilon_{yy}^{2}+\varepsilon_{zz}^{2})\nonumber \\
&+&2NV_{0}C_{xyxy}(\varepsilon_{xy}^{2}+\varepsilon_{yz}^{2}+\varepsilon_{zx}^{2}) \nonumber\\ 
&+&NV_{0}C_{xxyy}(\varepsilon_{xx}\varepsilon_{yy}+\varepsilon_{xx}\varepsilon_{zz}+\varepsilon_{yy}\varepsilon_{zz})
\end{eqnarray}

The uniform strain is independent compared with the vibronic distortions for the static JT case. 
To study the contribution of uniform strain, we should express the contribution of uniform strain in the same basis set as the study of JT effect. 

In the octahedral case, the symmetry should be $O_{h}$, where the Cartesian index, x, y, z, has the same symmetry of $T_{1u}$ element.
So the symmetrized basis set should be determined by the direct product $T_{1u}\otimes T_{1u}=A_{1}\oplus E\oplus\{T_{1}\}\oplus T_{2}$. 
From the Koster's group tables, we have the relation between the symmetrized strains and the general strains, which are also the elements of $\Upsilon$ matrix defined in Eq. \ref{eq:SymDefStrain}, as 
\begin{align}
\varepsilon_{xx} & =\frac{1}{\sqrt{3}}\varepsilon_{A_{1}}-\frac{1}{\sqrt{6}}\varepsilon_{E(\theta)}+\frac{1}{\sqrt{2}}\varepsilon_{E(\epsilon)}\nonumber \\
\varepsilon_{yy} & =\frac{1}{\sqrt{3}}\varepsilon_{A_{1}}-\frac{1}{\sqrt{6}}\varepsilon_{E(\theta)}-\frac{1}{\sqrt{2}}\varepsilon_{E(\epsilon)}\nonumber \\
\varepsilon_{xx} & =\frac{1}{\sqrt{3}}\varepsilon_{A_{1}}+\frac{\sqrt{2}}{\sqrt{3}}\varepsilon_{E(\theta)}\label{eq:SymUstrain}\\
\varepsilon_{yz} & =\varepsilon_{T_{2}(\xi)}\nonumber \\
\varepsilon_{zx} & =\varepsilon_{T_{2}(\eta)}\nonumber \\
\varepsilon_{xy} & =\varepsilon_{T_{2}(\zeta)}\nonumber 
\end{align}

Substituting Eq. \ref{eq:SymUstrain} into Eq. \ref{eq:UusLandau},
we have
\begin{eqnarray}
U_{us} &=&\frac{NV_{0}}{2}\left[\left(C_{xxxx}+2C_{xxyy}\right)e_{A_{1}}^{2}\right. \nonumber\\
&+&\left(C_{xxxx}-C_{xxyy}\right)e_{E(\theta)}^{2}+\left(C_{xxxx}-C_{xxyy}\right)\varepsilon_{E(\epsilon)}^{2} \nonumber \\
&+&\left.4C_{xyxy}\left(\varepsilon_{T_{2}(\xi)}^{2}+\varepsilon_{T_{2}(\eta)}^{2}+\varepsilon_{T_{2}(\zeta)}^{2}\right)\right]\label{eq:expressionCsymmtric}
\end{eqnarray}

Comparing with the form in Eq. \ref{eq:symmetrizedUniformStrain}, we have the correspondence of general elastic modulus and the symmetrized
elastic modulus as 
\begin{align}
C_{A_{1}} & =C_{xxxx}+2C_{xxyy}\nonumber \\
C_{E(\theta)} & =C_{E(\epsilon)}=C_{xxxx}-C_{xxyy}\\
C_{T_{2}(\xi)} & =C_{T_{2}(\eta)}=C_{T_{2}(\zeta)}=4C_{xyxy}\nonumber 
\end{align}

\section{$K$ matrix for isolated JT impurities}
\label{sec:appendix_IsolatedJTimp}
This is to check the trend of \emph{$K$} matrix elements with respect to the distance between two JT centers, for which there are three situations: 1). only considering the isolated one JT center, 2). considering JT pairs separated with a distance $\mathbf{n}$.
The following derivations and calculations were done with the example of Ba$_{2}$NaOsO$_{6}$. 
In order to have a better comparison of the strength of elastic interactions, $K$ matrix is transformed based
on the dimensionless JT modes: $K^{dl}=K/(\omega\times M_{am})$.

\subsection{Case 1: 1body JT center}

In this situation, JT centers are treated as impurities, which means,
there are only the two JT centers in the crystals. Thus, the K metrix
will be calculated directly from $\Xi$ matrix. 

For on-site situation, the $\Xi$ matrix considered should be a $2\times2$
matrix as

\begin{equation}
\Xi_{1body}=\left(\begin{array}{cc}
\Xi_{1E_{\theta}}^{1E_{\theta}} & \Xi_{1E_{\epsilon}}^{1E_{\theta}}\\
\Xi_{1E_{\theta}}^{1E_{\epsilon}} & \Xi_{1E_{\epsilon}}^{1E_{\epsilon}}
\end{array}\right)
\end{equation}

where, due to symmetry, $\Xi_{1E_{\theta}}^{1E_{\theta}}=\Xi_{1E_{\epsilon}}^{1E_{\epsilon}}=\Xi_{0}$,
$\Xi_{1E_{\epsilon}}^{1E_{\theta}}=\Xi_{1E_{\theta}}^{1E_{\epsilon}}=0$.

The corresponding $K_{1body}$ matrix is determined by directly inverse
the $\Xi_{1body}$ matrix, 

\begin{equation}
K_{1body}=\left(\begin{array}{cc}
K_{1E_{\theta}}^{1E_{\theta}} & K_{1E_{\epsilon}}^{1E_{\theta}}\\
K_{1E_{\theta}}^{1E_{\epsilon}} & K_{1E_{\epsilon}}^{1E_{\epsilon}}
\end{array}\right)=\left(\begin{array}{cc}
\frac{1}{\Xi_{0}} & 0\\
0 & \frac{1}{\Xi_{0}}
\end{array}\right)
\end{equation}

and the corresponding frequencies $\omega_{1body}=\sqrt{1/(\Xi_{0}*M^{am})}$.

\subsection{Case 2: JT interacting pairs}
For the interactions involving two JT centers, the $\Xi$ matrix should
be a $4\times4$ matrix, consisting of two $2\times2$ matrices corresponding
to 2 JT centers, as

\begin{align}
\Xi_{2body}(\mathbf{R}) & =\left(\begin{array}{cccc}
\Xi_{1E_{\theta}}^{1E_{\theta}} & \Xi_{1E_{\epsilon}}^{1E_{\theta}} & \Xi_{2E_{\theta}}^{1E_{\theta}}(\mathbf{R}) & \Xi_{2E_{\epsilon}}^{1E_{\theta}}(\mathbf{R})\\
\Xi_{1E_{\theta}}^{1E_{\epsilon}} & \Xi_{1E_{\epsilon}}^{1E_{\epsilon}} & \Xi_{2E_{\theta}}^{1E_{\epsilon}}(\mathbf{R}) & \Xi_{2E_{\epsilon}}^{1E_{\epsilon}}(\mathbf{R})\\
\Xi_{1E_{\theta}}^{2E_{\theta}}(\mathbf{R}) & \Xi_{1E_{\epsilon}}^{2E_{\theta}}(\mathbf{R}) & \Xi_{2E_{\theta}}^{2E_{\theta}} & \Xi_{2E_{\epsilon}}^{2E_{\theta}}\\
\Xi_{1E_{\theta}}^{2E_{\epsilon}}(\mathbf{R}) & \Xi_{1E_{\epsilon}}^{2E_{\epsilon}}(\mathbf{R}) & \Xi_{2E_{\theta}}^{2E_{\epsilon}} & \Xi_{2E_{\epsilon}}^{2E_{\epsilon}}
\end{array}\right)\\
 & =\left(\begin{array}{cccc}
\Xi_{1E_{\theta}}^{1E_{\theta}} & 0 & \Xi_{2E_{\theta}}^{1E_{\theta}}(\mathbf{R}) & \Xi_{2E_{\epsilon}}^{1E_{\theta}}(\mathbf{R})\\
0 & \Xi_{1E_{\epsilon}}^{1E_{\epsilon}} & \Xi_{2E_{\theta}}^{1E_{\epsilon}}(\mathbf{R}) & \Xi_{2E_{\epsilon}}^{1E_{\epsilon}}(\mathbf{R})\\
\Xi_{1E_{\theta}}^{2E_{\theta}}(\mathbf{R}) & \Xi_{1E_{\epsilon}}^{2E_{\theta}}(\mathbf{R}) & \Xi_{1E_{\theta}}^{1E_{\theta}} & 0\\
\Xi_{1E_{\theta}}^{2E_{\epsilon}}(\mathbf{R}) & \Xi_{1E_{\epsilon}}^{2E_{\epsilon}}(\mathbf{R}) & 0 & \Xi_{1E_{\epsilon}}^{1E_{\epsilon}}
\end{array}\right)
\end{align}
, where $\mathbf{R}=\mathbf{n}-\mathbf{n}'$ is the vector between the two JT centers locating in $\mathbf{n}$, and $mathbf{n}'$, respectively. 

The corresponding $K_{2body}(\mathbf{R})$ matrix is determined by the direct inverse of $\Xi_{2body}(\mathbf{R})$:

\begin{align}
K_{2body}(\mathbf{R}) & =inv\left(\begin{array}{cccc}
\Xi_{1E_{\theta}}^{1E_{\theta}} & 0 & \Xi_{2E_{\theta}}^{1E_{\theta}}(\mathbf{R}) & \Xi_{2E_{\epsilon}}^{1E_{\theta}}(\mathbf{R})\\
0 & \Xi_{1E_{\epsilon}}^{1E_{\epsilon}} & \Xi_{2E_{\theta}}^{1E_{\epsilon}}(\mathbf{R}) & \Xi_{2E_{\epsilon}}^{1E_{\epsilon}}(\mathbf{R})\\
\Xi_{1E_{\theta}}^{2E_{\theta}}(\mathbf{R}) & \Xi_{1E_{\epsilon}}^{2E_{\theta}}(\mathbf{R}) & \Xi_{1E_{\theta}}^{1E_{\theta}} & 0\\
\Xi_{1E_{\theta}}^{2E_{\epsilon}}(\mathbf{R}) & \Xi_{1E_{\epsilon}}^{2E_{\epsilon}}(\mathbf{R}) & 0 & \Xi_{1E_{\epsilon}}^{1E_{\epsilon}}
\end{array}\right)\\
 & =\left(\begin{array}{cccc}
K_{1E_{\theta}}^{1E_{\theta}} & K_{1E_{\epsilon}}^{1E_{\theta}} & K_{2E_{\theta}}^{1E_{\theta}}(\mathbf{R}) & K_{2E_{\epsilon}}^{1E_{\theta}}(\mathbf{R})\\
K_{1E_{\theta}}^{1E_{\epsilon}} & K_{1E_{\epsilon}}^{1E_{\epsilon}} & K_{2E_{\theta}}^{1E_{\epsilon}}(\mathbf{R}) & K_{2E_{\epsilon}}^{1E_{\epsilon}}(\mathbf{R})\\
K_{1E_{\theta}}^{2E_{\theta}}(\mathbf{R}) & K_{1E_{\epsilon}}^{2E_{\theta}}(\mathbf{R}) & K_{2E_{\theta}}^{2E_{\theta}} & K_{2E_{\epsilon}}^{2E_{\theta}}\\
K_{1E_{\theta}}^{2E_{\epsilon}}(\mathbf{R}) & K_{1E_{\epsilon}}^{2E_{\epsilon}}(\mathbf{R}) & K_{2E_{\theta}}^{2E_{\epsilon}} & K_{2E_{\epsilon}}^{2E_{\epsilon}}
\end{array}\right).
\end{align}
Since the $2\times 2$ block of the onsite interaction might not be diagonal, a diagonalization procedure is required. from which the frequencies are determined from the $K_{1E_{\theta}}^{1E_{\theta}}(\mathbf{R}=\mathbf{0},diag)$:
$\omega_{crystal}=\sqrt{K_{1E_{\theta}}^{1E_{\theta}}(\mathbf{R}=\mathbf{0},diag)/M^{am}}$. 

\bibliography{PRB_DoubleProvskite.bib}



\pagebreak
\clearpage
\widetext
\begin{center}
	\textbf{\large Supplemental Materials: First-principles derivation of elastic interaction between Jahn-Teller centers in crystals via LGFs}
\end{center}
\date{\today}
\setcounter{equation}{0}
\setcounter{figure}{0}
\setcounter{table}{0}
\setcounter{page}{1}
\setcounter{section}{0}
\makeatletter
\renewcommand{\theequation}{S\arabic{equation}}
\renewcommand{\thefigure}{S\arabic{figure}}
\renewcommand{\thetable}{S\arabic{table}}
\renewcommand{\thesection}{S\arabic{section}}
\renewcommand{\bibnumfmt}[1]{[S#1]}
\renewcommand{\citenumfont}[1]{S#1}


Before showing the information in this document, one thing should be noted that for simplicity, `LiOs', `NaOs', `MgRe', `ZnRe', `YW', `TaCl', and `TaBr' represent for Ba$_2$LiOsO$_6$, Ba$_2$NaOsO$_6$, Ba$_2$MgReO$_6$, Ba$_2$ZnReO$_6$, Ba$_2$YWO$_6$, Cs$_2$TaCl$_6$, and Cs$_2$TaBr$_6$, respectively.

\begin{figure*}
	\includegraphics[scale=0.8]{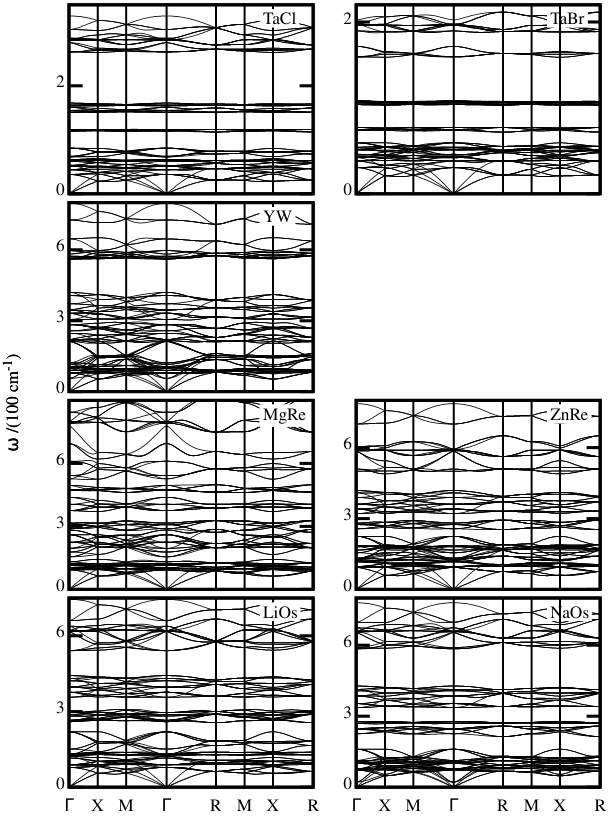}
	\caption{Phonon dispersions of double perovskites. 
	}
	\label{Fig:phonondispersion_all}
\end{figure*}

This document includes the following information:

The phonon dispersions of all the seven double perovskites studied in this paper are shown in Fig. \ref{Fig:phonondispersion_all}.

The position order of atoms in the convention unit cell is shown in Table. \ref{Table:Correspondence_atomicPosition}.

The $\Xi$ matrix from full phonons, for the inter site, the nearest, next nearest, next next nearest, next next next nearest, next next next next nearest sites, for all the double perovskites are shown in Table. \ref{tab:ECCs_summary_allori}.
The $\Xi$ matrix contributed from acoustic phonons, for the inter site, the nearest, next nearest, next next nearest, next next next nearest, next next next next nearest sites, for all the double perovskites are shown in Table. \ref{tab:ECCs_summary_allori}.

All the $\Xi$ matrix from the uniform strain are shown in Table. \ref{tab:ECCsUS_summary_all}.

The $K$ matrix of only $E_g$ JT active modes for the original, nearest, next nearest convention unit cells, from full phonons are shown in Table. \ref{tab:K_real_summary_allB11}.


\begin{table}
	\caption{$Fm\overline{3}m$ position correspondence of A$_{2}$B'BX$_{6}$, x should be provided to fully determine the structure.}
\label{Table:Correspondence_atomicPosition}
\begin{tabular}{|c|c|c|c|c|c|c|c|}
\hline 
P\_label & atom\_in & crystal position & P\_label & atom\_in & crystal position & atom\_in & crystal position\tabularnewline
\hline 
\hline 
\multirow{8}{*}{A(8c)} & 1 & (0.25, 0.25, 0.25) & \multirow{12}{*}{X(24e)} & 17 & (x,0.00,0.00) & 29 & (0.50+x,0.00,0.50)\tabularnewline
\cline{2-3} \cline{3-3} \cline{5-8} \cline{6-8} \cline{7-8} \cline{8-8} 
 & 2 & (0.75, 0.75, 0.75) &  & 18 & (1.00-x,0.00,0.00) & 30 & (0.50-x,0.00,0.50)\tabularnewline
\cline{2-3} \cline{3-3} \cline{5-8} \cline{6-8} \cline{7-8} \cline{8-8} 
 & 3 & (0.75, 0.75, 0.25) &  & 19 & (0.00,x,0.00) & 31 & (0.50,x,0.50)\tabularnewline
\cline{2-3} \cline{3-3} \cline{5-8} \cline{6-8} \cline{7-8} \cline{8-8} 
 & 4 & (0.25, 0.25, 0.75) &  & 20 & (0.00,1.00-x,0.00) & 32 & (0.50,1.00-x,0.50)\tabularnewline
\cline{2-3} \cline{3-3} \cline{5-8} \cline{6-8} \cline{7-8} \cline{8-8} 
 & 5 & (0.75, 0.25, 0.75) &  & 21 & (0.00,0.00,x) & 33 & (0.50,0.00,0.50+x)\tabularnewline
\cline{2-3} \cline{3-3} \cline{5-8} \cline{6-8} \cline{7-8} \cline{8-8} 
 & 6 & (0.25, 0.75, 0.25) &  & 22 & (0.00,0.00,1.00-x) & 34 & (0.50,0.00,0.50-x)\tabularnewline
\cline{2-3} \cline{3-3} \cline{5-8} \cline{6-8} \cline{7-8} \cline{8-8} 
 & 7 & (0.25, 0.75, 0.75) &  & 23 & (x,0.50,0.50) & 35 & (0.50+x,0.50,0.00)\tabularnewline
\cline{2-3} \cline{3-3} \cline{5-8} \cline{6-8} \cline{7-8} \cline{8-8} 
 & 8 & (0.75, 0.25, 0.25) &  & 24 & (1.00-x,0.50,0.50) & 36 & (0.50-x,0.50,0.00)\tabularnewline
\cline{1-3} \cline{2-3} \cline{3-3} \cline{5-8} \cline{6-8} \cline{7-8} \cline{8-8} 
\multirow{4}{*}{B'(4b)} & 9 & (0.50,0.50,0.50) &  & 25 & (0.00,0.50+x,0.50) & 37 & (0.50,0.50+x,0.00)\tabularnewline
\cline{2-3} \cline{3-3} \cline{5-8} \cline{6-8} \cline{7-8} \cline{8-8} 
 & 10 & (0.50,0.00,0.00) &  & 26 & (0.00,0.50-x,0.50) & 38 & (0.50,0.50-x,0.00)\tabularnewline
\cline{2-3} \cline{3-3} \cline{5-8} \cline{6-8} \cline{7-8} \cline{8-8} 
 & 11 & (0.00,0.50,0.00) &  & 27 & (0.00,0.50,0.50+x) & 39 & (0.50,0.50,x)\tabularnewline
\cline{2-3} \cline{3-3} \cline{5-8} \cline{6-8} \cline{7-8} \cline{8-8} 
 & 12 & (0.00,0.00,0.50) &  & 28 & (0.00,0.50,0.50-x) & 40 & (0.50,0.50,1.00-x)\tabularnewline
\hline 
\multirow{4}{*}{B(4a)} & 13 & (0.00,0.00,0.00) &  &  &  &  & \tabularnewline
\cline{2-8} \cline{3-8} \cline{4-8} \cline{5-8} \cline{6-8} \cline{7-8} \cline{8-8} 
 & 14 & (0.00,0.50,0.50) &  &  &  &  & \tabularnewline
\cline{2-8} \cline{3-8} \cline{4-8} \cline{5-8} \cline{6-8} \cline{7-8} \cline{8-8} 
 & 15 & (0.50,0.00,0.50) &  &  &  &  & \tabularnewline
\cline{2-8} \cline{3-8} \cline{4-8} \cline{5-8} \cline{6-8} \cline{7-8} \cline{8-8} 
 & 16 & (0.50,0.50,0.00) &  &  &  &  & \tabularnewline
\hline 
\end{tabular}
\end{table}

\input{./Eccs_summary_Latex_5_ori}
\input{./EccsUS_summary_Latex_5}


\input{./B11Kappa_real_summary_Latex_8}



The $K$ matrix from the uniform strain is summarized in Table. \ref{tab:K_real_summary_allUS}.
\input{./USKappa_real_summary_Latex_8}

The $K$ matrix from IFCs are shown in Table. \ref{tab:K_real_summary_allIFCKappa}.
\input{./IFCKappaKappa_real_summary_Latex_8}

\end{document}

%% file: Eccs_summary_all_US
\begin{table}  
\caption{\label{tab:ECCs_summary_all_includingUS} 
$\Xi$ matrix for JT active centers in one convention unit cell. 
`PH' and `US' distinguish the parameters from phonons and uniform strain, respectively. 
From the left to right, and from the top to bottom, the matrix elements correspond to JTD\_2, JTD\_3, JTD\_4, JTD\_5 and JTD\_6, respectively. 
The unit is $bohr^2/Ha$. `NaOs' and `LiOs' are for 'Ba$_2$NaOsO$_6$' and 'Ba$_2$LiOsO$_6$', respectively. } 
\resizebox{0.5\textwidth}{!}{  
\begin{tabular}{|c|c|c|}
\hline  
& NaOs &  LiOs  \\
\hline
$\begin{array}{c}
PH \\
(JTC1,JTC1)
\end{array}$
 & 
$ \left( \begin{array}{ccccc}4.988 &0.000 &0.000 &0.000 &0.000 \\ 
0.000 &4.988 &0.000 &0.000 &0.000 \\ 
0.000 &0.000 &10.472 &0.000 &0.000 \\ 
0.000 &0.000 &0.000 &10.472 &0.000 \\ 
0.000 &0.000 &0.000 &0.000 &10.472 \\
\end{array}\right) $
&
$ \left( \begin{array}{ccccc}5.464 &0.000 &0.000 &0.000 &0.000 \\ 
0.000 &5.464 &0.000 &0.000 &0.000 \\ 
0.000 &0.000 &9.921 &0.000 &0.000 \\ 
0.000 &0.000 &0.000 &9.921 &0.000 \\ 
0.000 &0.000 &0.000 &0.000 &9.921 \\
\end{array}\right)$ \\
\hline
$\begin{array}{c}
PH \\
(JTC1,JTC2)
\end{array}$
 & 
$ \left( \begin{array}{ccccc} 0.065 & 0.035 & $-0.037$ & 0.000 & 0.000 \\
   0.035 & 0.025 & 0.065 & 0.000 & 0.000 \\
  $-0.037$&0.065 &$-0.200$&0.000 & 0.000 \\
   0.000 & 0.000 & 0.000 & 0.058 & 0.025 \\
   0.000 & 0.000 & 0.000 & 0.025 & 0.058 \\
\end{array}\right) $
&
$ \left( \begin{array}{ccccc} 0.075 & 0.063 & $-0.051$ & 0.000 & 0.000 \\
   0.063 & 0.002 & 0.088 & 0.000 & 0.000 \\
  $-0.051$&0.088 &$-0.204$&0.000& 0.000 \\
   0.000 & 0.000 & 0.000 & 0.065& 0.016 \\
   0.000 & 0.000 & 0.000 & 0.016& 0.065 \\
\end{array}\right) $ \\
\hline
$\begin{array}{c} 
US \\ 
(JTC1,JTC1)
\end{array}$
 &
$ \left( \begin{array}{ccccc}0.267 &0.000 &0.000 &0.000 &0.000 \\ 
0.000 &0.267 &0.000 &0.000 &0.000 \\ 
0.000 &0.000 &0.313 &0.000 &0.000 \\ 
0.000 &0.000 &0.000 &0.313 &0.000 \\ 
0.000 &0.000 &0.000 &0.000 &0.313 \\
\end{array}\right) $ 
&
$  \left( \begin{array}{ccccc}0.386 &0.000 &0.000 &0.000 &0.000 \\ 
0.000 &0.386 &0.000 &0.000 &0.000 \\ 
0.000 &0.000 &0.243 &0.000 &0.000 \\ 
0.000 &0.000 &0.000 &0.243 &0.000 \\ 
0.000 &0.000 &0.000 &0.000 &0.243 \\
\end{array}\right)$ \\
\hline
$\begin{array}{c}
US \\
(JTC1,JTC2)
\end{array}$
&
$ \left( \begin{array}{ccccc}0.033 &-0.081 &0.000 &0.000 &0.000 \\ 
-0.081 &0.127 &0.000 &0.000 &0.000 \\ 
0.000 &0.000 &-0.016 &0.000 &0.000 \\ 
0.000 &0.000 &0.000 &0.148 &0.000 \\ 
0.000 &0.000 &0.000 &0.000 &0.148 \\
\end{array}\right) $
&
$ \left( \begin{array}{ccccc}0.047 &-0.117 &0.000 &0.000 &0.000 \\ 
-0.117 &0.182 &0.000 &0.000 &0.000 \\ 
0.000 &0.000 &-0.013 &0.000 &0.000 \\ 
0.000 &0.000 &0.000 &0.115 &0.000 \\ 
0.000 &0.000 &0.000 &0.000 &0.115 \\ 
\end{array}\right) $\\
\hline
\end{tabular}
}
\end{table}

%% file: B11Kappa_real_summary_Latex_8_Eg
\begin{table*}  
\caption{\label{tab:K_real_summary_EgB11} 
Contribution from uniform strainThe distance dependent $K ^{\mu \gamma}_{\mu' \gamma'} (\bm{R})$ matrix in real space.The matrix correspondd to $ \left( \begin{array}{cc} K ^{E_{g\theta}}_{E_{g\theta}} &K ^{E_{g\theta}}_{E_{g\epsilon}} \\ K ^{E_{g\epsilon}}_{E_{g\theta}} & K ^{E_{g\epsilon}}_{E_{g\epsilon}}  \\ \end{array}\right) $. $R$ is the crystal constant. LGF, 2body, 1body and 1body(IFC) represent the data calculated with the method proposed in this paper, two-body coupling impurity, and one-body coupling, and calculated directly with interatomic force constants from first-principles calculations, respectively. The unit is meV.} 
\resizebox{\textwidth}{!}{  
\begin{tabular}{|c|c|c|c|c|c|c|c|}\hline  
  &      &  $0 R$   &  $\sqrt{2}/2 R$   &  $1 R$   &  $\sqrt{6}/2 R$   &  $2\sqrt{2}/2 R$   &  $\sqrt{10}/2 R$  \\   
\hline  
\multirow{2}{*}{ NaOs } & LGFs & $\left( \begin{array}{cc} 71.176 & 0.000 \\
0.000 & 71.176 \\
  \end{array}\right)$ &$\left( \begin{array}{cc} -0.903 & -0.487 \\
-0.487 & -0.340 \\
  \end{array}\right)$ &$\left( \begin{array}{cc} -0.338 & -0.866 \\
-0.866 & 0.662 \\
  \end{array}\right)$ &$\left( \begin{array}{cc} -0.099 & -0.064 \\
-0.064 & -0.026 \\
  \end{array}\right)$ &$\left( \begin{array}{cc} 0.114 & 0.000 \\
0.000 & -0.241 \\
  \end{array}\right)$ &$\left( \begin{array}{cc} -0.902 & 0.505 \\
0.492 & -0.334 \\
  \end{array}\right)$   \\   \cline{2-8} & 2body & $\left( \begin{array}{cc} 71.358 & 0.009 \\
0.009 & 71.347 \\
  \end{array}\right)$ &$\left( \begin{array}{cc} -0.932 & -0.496 \\
-0.496 & -0.360 \\
  \end{array}\right)$ &$\left( \begin{array}{cc} -1.844 & -0.858 \\
-0.858 & -0.853 \\
  \end{array}\right)$ &$\left( \begin{array}{cc} -0.148 & -0.066 \\
-0.066 & -0.072 \\
  \end{array}\right)$ &$\left( \begin{array}{cc} 0.081 & -0.000 \\
0.000 & -0.341 \\
  \end{array}\right)$ &$\left( \begin{array}{cc} -0.932 & 0.496 \\
0.496 & -0.360 \\
  \end{array}\right)$   \\   \cline{2-8} & 1body & $\left( \begin{array}{cc} 71.342 & 0.000 \\
0.000 & 71.342 \\
  \end{array}\right)$ &- &- &- &- &-   \\   \cline{2-8} & 1body (IFC) & $\left( \begin{array}{cc} 75.840 & 0.000 \\
0.000 & 75.840 \\
  \end{array}\right)$ &$\left( \begin{array}{cc} -1.059 & -1.040 \\
-1.040 & 0.143 \\
  \end{array}\right)$ &- &- &- &- \\  
\hline  
\multirow{2}{*}{ LiOs } & LGFs & $\left( \begin{array}{cc} 68.005 & 0.000 \\
0.000 & 68.005 \\
  \end{array}\right)$ &$\left( \begin{array}{cc} -0.901 & -0.759 \\
-0.759 & -0.024 \\
  \end{array}\right)$ &$\left( \begin{array}{cc} -0.388 & -0.384 \\
-0.384 & 0.055 \\
  \end{array}\right)$ &$\left( \begin{array}{cc} -0.126 & -0.074 \\
-0.074 & -0.041 \\
  \end{array}\right)$ &$\left( \begin{array}{cc} 0.183 & -0.000 \\
-0.000 & -0.046 \\
  \end{array}\right)$ &$\left( \begin{array}{cc} -0.907 & 0.780 \\
0.755 & -0.021 \\
  \end{array}\right)$   \\   \cline{2-8} & 2body & $\left( \begin{array}{cc} 68.185 & 0.011 \\
0.011 & 68.172 \\
  \end{array}\right)$ &$\left( \begin{array}{cc} -0.937 & -0.791 \\
-0.791 & -0.023 \\
  \end{array}\right)$ &$\left( \begin{array}{cc} -1.858 & -0.370 \\
-0.370 & -1.431 \\
  \end{array}\right)$ &$\left( \begin{array}{cc} -0.146 & -0.076 \\
-0.076 & -0.058 \\
  \end{array}\right)$ &$\left( \begin{array}{cc} 0.115 & -0.000 \\
0.000 & -0.197 \\
  \end{array}\right)$ &$\left( \begin{array}{cc} -0.937 & 0.791 \\
0.791 & -0.023 \\
  \end{array}\right)$   \\   \cline{2-8} & 1body & $\left( \begin{array}{cc} 68.164 & 0.000 \\
0.000 & 68.164 \\
  \end{array}\right)$ &- &- &- &- &-   \\   \cline{2-8} & 1body (IFC) & $\left( \begin{array}{cc} 71.103 & 0.000 \\
-0.000 & 71.103 \\
  \end{array}\right)$ &$\left( \begin{array}{cc} -1.366 & -1.249 \\
-1.249 & 0.077 \\
  \end{array}\right)$ &- &- &- &- \\  
\hline  
\end{tabular}  
}  
\end{table*}  

%% file: US2Kappa_real_summary_Latex_8
\begin{table}  
\caption{\label{tab:K_real_summary_allUS2} 
Contribution from uniform strain: $K ^{\mu \gamma}_{\mu' \gamma'} $ matrix in real space. The each 2$\times$2 block correspondd to $ \left( \begin{array}{cc} K ^{E_{g\theta}}_{E_{g\theta}} &K ^{E_{g\theta}}_{E_{g\epsilon}} \\ K ^{E_{g\epsilon}}_{E_{g\theta}} & K ^{E_{g\epsilon}}_{E_{g\epsilon}}  \\ \end{array}\right) $ of one JT center. The unit is meV.} 
\resizebox{0.5\textwidth}{!}{  
\begin{tabular}{|c|c|}\hline  
&  onsite  \\   
\hline  
NaOs & $ \left( \begin{array}{cccccccc}-14.040 &0.000 &-0.438 &4.712 &-0.438 &-4.712 &-8.599 &-0.000 \\ 
0.000 &-14.040 &4.712 &-5.878 &-4.712 &-5.878 &-0.000 &2.283 \\ 
-0.438 &4.712 &-2.061 &2.927 &1.102 &1.019 &0.664 &-0.766 \\ 
4.712 &-5.878 &2.927 &-5.441 &-1.019 &0.518 &1.271 &0.956 \\ 
-0.438 &-4.712 &1.102 &-1.019 &-2.061 &-2.927 &0.664 &0.766 \\ 
-4.712 &-5.878 &1.019 &0.518 &-2.927 &-5.441 &-1.271 &0.956 \\ 
-8.599 &-0.000 &0.664 &1.271 &0.664 &-1.271 &-7.131 &0.000 \\ 
-0.000 &2.283 &-0.766 &0.956 &0.766 &0.956 &0.000 &-0.371 \\ 
\end{array}\right) $\\ 
\hline  
LiOs & $ \left( \begin{array}{cccccccc}-15.532 &0.000 &-0.296 &5.283 &-0.283 &-5.283 &-9.481 &-0.001 \\ 
-0.000 &-15.532 &5.285 &-6.399 &-5.302 &-6.414 &0.002 &2.781 \\ 
-0.285 &5.309 &-2.184 &2.928 &1.437 &1.452 &0.567 &-0.951 \\ 
5.311 &-6.415 &2.928 &-5.565 &-1.452 &0.272 &1.959 &1.149 \\ 
-0.298 &-5.292 &1.421 &-1.432 &-2.184 &-2.933 &0.563 &0.947 \\ 
-5.311 &-6.399 &1.431 &0.287 &-2.923 &-5.565 &-1.957 &1.146 \\ 
-9.449 &-0.002 &0.567 &1.920 &0.571 &-1.923 &-7.255 &0.000 \\ 
0.001 &2.754 &-0.937 &1.134 &0.940 &1.138 &-0.000 &-0.493 \\ 
\end{array}\right) $\\ 
\hline  
\end{tabular}  
}  
\end{table}

%% file: B11Summary_freq_Eg
\begin{table}  
\caption{\label{Table:Omega_K} 
The vibration frequency ($cm^{-1}$) of E$_g$ JT active modes, calculated by the diagonalizing of the $K$ matrix. `LiOs', `NaOs', `MgRe', `ZnRe', `YW', `TaCl', and `TaBr' represent for Ba$_2$NaOsO$_6$, Ba$_2$LiOsO$_6$, Ba$_2$MgReO$_6$, Ba$_2$ZnReO$_6$, Cs$_2$TaCl$_6$, Cs$_2$TaBr$_6$,  and Ba$_2$YWO$_6$. LGF, 2body, 1body and 1body(IFC) represent the data calculated with the method proposed in this paper, two-body coupling impurity, and one-body couplingand calculated directly with interatomic force constants from first-principles calculations, respectively.} 
\begin{tabular}{|c|c|c|c|c|}\hline  
& LGF & 2body & 1body & 1body(IFC) \\ 
\hline 
NaOs & 577.523 & 575.412 & 575.412 & 611.687 \\
\hline 
LiOs & 551.796 & 549.781 & 549.792 & 573.481 \\
\hline 
MgRe & 543.886 & 541.710 & 541.710 & 577.305 \\
\hline 
ZnRe & 526.965 & 524.773 & 524.799 & 563.498 \\
\hline 
TaCl & 266.012 & 264.996 & 264.996 & 268.737 \\
\hline 
TaBr & 243.006 & 242.074 & 242.074 & 245.852 \\
\hline 
YW & 560.837 & 557.397 & 557.418 & 604.508 \\
\hline 
\end{tabular}  
\end{table}  

%% file: Eccs_summary_Latex_5_ori
\begin{sidewaystable*}[b]  
\caption{\label{tab:ECCs_summary_allori} 
$\Xi$ matrix for both intra and inter JT active centers. The inter site interaction is labeled as 1N, 2N, 3N, 4N, 5N, for the nearest, next nearest, next next nearest, next next next nearest, next next next next nearest sites, respectively. JTC1, JTC2, JTC3 are the label for the JT active centers, as defined in Table. \ref{Table:coordinates_SAJTmodes_Oct}. From the left to right, and from the top to bottom, the matrix elements correspond to JTD\_2, JTD\_3, JTD\_4, JTD\_5 and JTD\_6, respectively. The unit is $m_e bohr^2/Ry$'} 
\resizebox{\textwidth}{!}{  
\begin{tabular}{|c|c|c|c|c|c|c|}\hline  
&  Intra &  1N &  2N &  3N &  4N &  5N  \\   
\hline  
& (0,0,0) (JTC1,JTC1) & (0,0,0) (JTC1,JTC2) & (0,1,0) (JTC1,JTC1) & (0,1,0) (JTC1,JTC3) & (0,1,1) (JTC1,JTC1) & (0,1,0) (JTC1,JTC2) \\
\hline  
NaOs & $ \left( \begin{array}{ccccc}4.988 &0.000 &0.000 &0.000 &0.000 \\ 
0.000 &4.988 &0.000 &0.000 &0.000 \\ 
0.000 &0.000 &10.472 &0.000 &0.000 \\ 
0.000 &0.000 &0.000 &10.472 &0.000 \\ 
0.000 &0.000 &0.000 &0.000 &10.472 \\ 
\end{array}\right) $&$ \left( \begin{array}{ccccc}0.065 &0.035 &-0.037 &0.000 &0.000 \\ 
0.035 &0.025 &0.065 &0.000 &0.000 \\ 
-0.037 &0.065 &-0.200 &0.000 &0.000 \\ 
0.000 &0.000 &0.000 &0.058 &0.025 \\ 
0.000 &0.000 &0.000 &0.025 &0.058 \\ 
\end{array}\right) $&$ \left( \begin{array}{ccccc}0.129 &-0.060 &0.000 &0.000 &0.000 \\ 
-0.060 &0.060 &0.000 &0.000 &0.000 \\ 
0.000 &0.000 &0.338 &0.000 &0.000 \\ 
0.000 &0.000 &0.000 &0.348 &0.000 \\ 
0.000 &0.000 &0.000 &0.000 &0.338 \\ 
\end{array}\right) $&$ \left( \begin{array}{ccccc}0.010 &-0.005 &-0.008 &-0.001 &0.001 \\ 
-0.005 &0.005 &-0.003 &-0.001 &-0.008 \\ 
-0.008 &-0.003 &0.009 &-0.000 &-0.013 \\ 
-0.001 &-0.001 &-0.000 &-0.003 &-0.000 \\ 
0.001 &-0.008 &-0.013 &-0.000 &0.009 \\ 
\end{array}\right) $&$ \left( \begin{array}{ccccc}0.016 &0.013 &-0.010 &0.000 &0.000 \\ 
0.013 &0.002 &0.017 &0.000 &0.000 \\ 
-0.010 &0.017 &-0.020 &0.000 &0.000 \\ 
0.000 &0.000 &0.000 &0.010 &0.001 \\ 
0.000 &0.000 &0.000 &0.001 &0.010 \\ 
\end{array}\right) $&$ \left( \begin{array}{ccccc}0.001 &-0.008 &0.001 &0.000 &0.000 \\ 
-0.008 &-0.016 &0.014 &0.000 &0.000 \\ 
0.001 &0.014 &-0.001 &0.000 &0.000 \\ 
0.000 &0.000 &0.000 &0.001 &0.000 \\ 
0.000 &0.000 &0.000 &0.000 &0.003 \\ 
\end{array}\right) $\\ 
\hline  
LiOs & $ \left( \begin{array}{ccccc}5.464 &0.000 &0.000 &0.000 &0.000 \\ 
0.000 &5.464 &0.000 &0.000 &0.000 \\ 
0.000 &0.000 &9.921 &0.000 &0.000 \\ 
0.000 &0.000 &0.000 &9.921 &0.000 \\ 
0.000 &0.000 &0.000 &0.000 &9.921 \\ 
\end{array}\right) $&$ \left( \begin{array}{ccccc}0.075 &0.063 &-0.051 &0.000 &0.000 \\ 
0.063 &0.002 &0.088 &0.000 &0.000 \\ 
-0.051 &0.088 &-0.204 &0.000 &0.000 \\ 
0.000 &0.000 &0.000 &0.065 &0.016 \\ 
0.000 &0.000 &0.000 &0.016 &0.065 \\ 
\end{array}\right) $&$ \left( \begin{array}{ccccc}0.149 &-0.030 &0.000 &0.000 &0.000 \\ 
-0.030 &0.115 &0.000 &0.000 &0.000 \\ 
0.000 &0.000 &0.308 &0.000 &0.000 \\ 
0.000 &0.000 &0.000 &0.335 &0.000 \\ 
0.000 &0.000 &0.000 &0.000 &0.308 \\ 
\end{array}\right) $&$ \left( \begin{array}{ccccc}0.012 &-0.006 &-0.011 &0.000 &0.004 \\ 
-0.006 &0.005 &-0.002 &0.000 &-0.010 \\ 
-0.011 &-0.002 &0.007 &0.003 &-0.017 \\ 
0.000 &0.000 &0.002 &-0.001 &0.002 \\ 
0.003 &-0.011 &-0.017 &0.003 &0.007 \\ 
\end{array}\right) $&$ \left( \begin{array}{ccccc}0.010 &0.011 &-0.012 &0.000 &0.000 \\ 
0.011 &-0.003 &0.021 &0.000 &0.000 \\ 
-0.012 &0.021 &-0.020 &0.000 &0.000 \\ 
0.000 &0.000 &0.000 &0.011 &0.003 \\ 
0.000 &0.000 &0.000 &0.003 &0.011 \\ 
\end{array}\right) $&$ \left( \begin{array}{ccccc}0.000 &-0.002 &-0.002 &0.000 &0.000 \\ 
-0.003 &-0.005 &0.011 &0.000 &0.000 \\ 
-0.002 &0.011 &-0.003 &0.000 &0.000 \\ 
0.000 &0.000 &0.000 &0.002 &-0.000 \\ 
0.000 &0.000 &0.000 &-0.000 &0.004 \\ 
\end{array}\right) $\\ 
\hline  
MgRe & $ \left( \begin{array}{ccccc}5.628 &0.000 &0.000 &0.000 &0.000 \\ 
0.000 &5.628 &0.000 &0.000 &0.000 \\ 
0.000 &0.000 &11.029 &0.000 &0.000 \\ 
0.000 &0.000 &0.000 &11.029 &0.000 \\ 
0.000 &0.000 &0.000 &0.000 &11.029 \\ 
\end{array}\right) $&$ \left( \begin{array}{ccccc}0.156 &0.090 &-0.099 &0.000 &0.000 \\ 
0.090 &0.052 &0.172 &0.000 &0.000 \\ 
-0.099 &0.172 &-0.322 &0.000 &0.000 \\ 
0.000 &0.000 &0.000 &0.141 &0.022 \\ 
0.000 &0.000 &0.000 &0.022 &0.141 \\ 
\end{array}\right) $&$ \left( \begin{array}{ccccc}0.135 &-0.044 &0.000 &0.000 &0.000 \\ 
-0.044 &0.084 &0.000 &0.000 &0.000 \\ 
0.000 &0.000 &0.342 &0.000 &0.000 \\ 
0.000 &0.000 &0.000 &0.369 &0.000 \\ 
0.000 &0.000 &0.000 &0.000 &0.342 \\ 
\end{array}\right) $&$ \left( \begin{array}{ccccc}0.022 &-0.013 &-0.018 &0.003 &-0.000 \\ 
-0.013 &0.007 &-0.011 &0.005 &-0.021 \\ 
-0.018 &-0.011 &0.007 &0.008 &-0.032 \\ 
0.003 &0.005 &0.008 &0.010 &0.008 \\ 
-0.000 &-0.021 &-0.032 &0.008 &0.007 \\ 
\end{array}\right) $&$ \left( \begin{array}{ccccc}0.013 &0.021 &-0.022 &0.000 &0.000 \\ 
0.021 &-0.011 &0.039 &0.000 &0.000 \\ 
-0.022 &0.039 &-0.042 &0.000 &0.000 \\ 
0.000 &0.000 &0.000 &0.015 &0.003 \\ 
0.000 &0.000 &0.000 &0.003 &0.015 \\ 
\end{array}\right) $&$ \left( \begin{array}{ccccc}0.001 &-0.009 &-0.001 &0.000 &0.000 \\ 
-0.009 &-0.015 &0.025 &0.000 &0.000 \\ 
-0.001 &0.025 &-0.005 &0.000 &0.000 \\ 
0.000 &0.000 &0.000 &0.005 &0.000 \\ 
0.000 &0.000 &0.000 &0.000 &0.009 \\ 
\end{array}\right) $\\ 
\hline  
ZnRe & $ \left( \begin{array}{ccccc}5.997 &0.000 &0.000 &0.000 &0.000 \\ 
0.000 &5.997 &0.000 &0.000 &0.000 \\ 
0.000 &0.000 &11.913 &0.000 &0.000 \\ 
0.000 &0.000 &0.000 &11.913 &0.000 \\ 
0.000 &0.000 &0.000 &0.000 &11.913 \\ 
\end{array}\right) $&$ \left( \begin{array}{ccccc}0.154 &0.124 &-0.096 &0.000 &0.000 \\ 
0.124 &0.012 &0.166 &0.000 &0.000 \\ 
-0.096 &0.166 &-0.333 &0.000 &0.000 \\ 
0.000 &0.000 &0.000 &0.121 &0.036 \\ 
0.000 &0.000 &0.000 &0.036 &0.121 \\ 
\end{array}\right) $&$ \left( \begin{array}{ccccc}0.137 &-0.011 &0.000 &0.000 &0.000 \\ 
-0.011 &0.124 &0.000 &0.000 &0.000 \\ 
0.000 &0.000 &0.369 &0.000 &0.000 \\ 
0.000 &0.000 &0.000 &0.411 &0.000 \\ 
0.000 &0.000 &0.000 &0.000 &0.369 \\ 
\end{array}\right) $&$ \left( \begin{array}{ccccc}0.030 &-0.020 &-0.024 &0.001 &0.008 \\ 
-0.020 &0.007 &-0.005 &0.001 &-0.023 \\ 
-0.024 &-0.005 &0.011 &0.004 &-0.031 \\ 
0.001 &0.001 &0.004 &-0.000 &0.004 \\ 
0.008 &-0.023 &-0.031 &0.004 &0.011 \\ 
\end{array}\right) $&$ \left( \begin{array}{ccccc}0.010 &0.027 &-0.023 &0.000 &0.000 \\ 
0.027 &-0.021 &0.039 &0.000 &0.000 \\ 
-0.023 &0.039 &-0.035 &0.000 &0.000 \\ 
0.000 &0.000 &0.000 &0.011 &0.006 \\ 
0.000 &0.000 &0.000 &0.006 &0.011 \\ 
\end{array}\right) $&$ \left( \begin{array}{ccccc}-0.005 &-0.005 &-0.005 &0.000 &0.000 \\ 
-0.005 &-0.004 &0.016 &0.000 &0.000 \\ 
-0.005 &0.016 &-0.004 &0.000 &0.000 \\ 
0.000 &0.000 &0.000 &0.004 &0.002 \\ 
0.000 &0.000 &0.000 &0.002 &0.008 \\ 
\end{array}\right) $\\ 
\hline  
TaCl & $ \left( \begin{array}{ccccc}10.614 &0.000 &0.000 &0.000 &0.000 \\ 
0.000 &10.614 &0.000 &0.000 &0.000 \\ 
0.000 &0.000 &29.407 &0.000 &0.000 \\ 
0.000 &0.000 &0.000 &29.407 &0.000 \\ 
0.000 &0.000 &0.000 &0.000 &29.407 \\ 
\end{array}\right) $&$ \left( \begin{array}{ccccc}0.085 &0.051 &-0.085 &0.000 &0.000 \\ 
0.051 &0.026 &0.147 &0.000 &0.000 \\ 
-0.085 &0.147 &-0.014 &0.000 &0.000 \\ 
0.000 &0.000 &0.000 &0.331 &0.210 \\ 
0.000 &0.000 &0.000 &0.210 &0.331 \\ 
\end{array}\right) $&$ \left( \begin{array}{ccccc}0.326 &-0.024 &0.000 &0.000 &0.000 \\ 
-0.024 &0.298 &0.000 &0.000 &0.000 \\ 
0.000 &0.000 &0.954 &0.000 &0.000 \\ 
0.000 &0.000 &0.000 &0.915 &0.000 \\ 
0.000 &0.000 &0.000 &0.000 &0.954 \\ 
\end{array}\right) $&$ \left( \begin{array}{ccccc}0.011 &-0.009 &-0.017 &0.005 &0.001 \\ 
-0.009 &0.000 &-0.009 &0.008 &-0.019 \\ 
-0.017 &-0.009 &0.030 &0.003 &-0.019 \\ 
0.005 &0.008 &0.003 &-0.004 &0.003 \\ 
0.001 &-0.019 &-0.019 &0.003 &0.030 \\ 
\end{array}\right) $&$ \left( \begin{array}{ccccc}0.013 &0.011 &-0.015 &0.000 &0.000 \\ 
0.011 &0.001 &0.025 &0.000 &0.000 \\ 
-0.015 &0.025 &-0.011 &0.000 &0.000 \\ 
0.000 &0.000 &0.000 &0.037 &0.011 \\ 
0.000 &0.000 &0.000 &0.011 &0.037 \\ 
\end{array}\right) $&$ \left( \begin{array}{ccccc}0.001 &-0.004 &-0.002 &0.000 &0.000 \\ 
-0.004 &-0.009 &0.017 &0.000 &0.000 \\ 
-0.002 &0.017 &-0.001 &0.000 &0.000 \\ 
0.000 &0.000 &0.000 &0.000 &0.014 \\ 
0.000 &0.000 &0.000 &0.014 &0.006 \\ 
\end{array}\right) $\\ 
\hline  
TaBr & $ \left( \begin{array}{ccccc}12.719 &0.000 &0.000 &0.000 &0.000 \\ 
0.000 &12.719 &0.000 &0.000 &0.000 \\ 
0.000 &0.000 &31.722 &0.000 &0.000 \\ 
0.000 &0.000 &0.000 &31.722 &0.000 \\ 
0.000 &0.000 &0.000 &0.000 &31.722 \\ 
\end{array}\right) $&$ \left( \begin{array}{ccccc}0.120 &0.075 &-0.105 &0.000 &0.000 \\ 
0.075 &0.033 &0.181 &0.000 &0.000 \\ 
-0.105 &0.181 &-0.014 &0.000 &0.000 \\ 
0.000 &0.000 &0.000 &0.382 &0.217 \\ 
0.000 &0.000 &0.000 &0.217 &0.382 \\ 
\end{array}\right) $&$ \left( \begin{array}{ccccc}0.392 &-0.035 &0.000 &0.000 &0.000 \\ 
-0.035 &0.352 &0.000 &0.000 &0.000 \\ 
0.000 &0.000 &1.037 &0.000 &0.000 \\ 
0.000 &0.000 &0.000 &1.000 &0.000 \\ 
0.000 &0.000 &0.000 &0.000 &1.037 \\ 
\end{array}\right) $&$ \left( \begin{array}{ccccc}0.015 &-0.012 &-0.020 &0.007 &-0.001 \\ 
-0.012 &0.001 &-0.012 &0.013 &-0.024 \\ 
-0.020 &-0.012 &0.030 &0.006 &-0.025 \\ 
0.007 &0.013 &0.006 &-0.002 &0.006 \\ 
-0.001 &-0.024 &-0.025 &0.006 &0.030 \\ 
\end{array}\right) $&$ \left( \begin{array}{ccccc}0.018 &0.013 &-0.018 &0.000 &0.000 \\ 
0.013 &0.002 &0.031 &0.000 &0.000 \\ 
-0.018 &0.031 &-0.012 &0.000 &0.000 \\ 
0.000 &0.000 &0.000 &0.046 &0.011 \\ 
0.000 &0.000 &0.000 &0.011 &0.046 \\ 
\end{array}\right) $&$ \left( \begin{array}{ccccc}0.002 &-0.006 &-0.003 &0.000 &0.000 \\ 
-0.006 &-0.011 &0.022 &0.000 &0.000 \\ 
-0.003 &0.022 &-0.001 &0.000 &0.000 \\ 
0.000 &0.000 &0.000 &0.001 &0.014 \\ 
0.000 &0.000 &0.000 &0.014 &0.007 \\ 
\end{array}\right) $\\ 
\hline  
YW & $ \left( \begin{array}{ccccc}5.316 &0.000 &0.000 &0.000 &0.000 \\ 
0.000 &5.316 &0.000 &0.000 &0.000 \\ 
0.000 &0.000 &13.385 &0.000 &0.000 \\ 
0.000 &0.000 &0.000 &13.385 &0.000 \\ 
0.000 &0.000 &0.000 &0.000 &13.385 \\ 
\end{array}\right) $&$ \left( \begin{array}{ccccc}0.148 &0.059 &-0.085 &0.000 &0.000 \\ 
0.059 &0.080 &0.147 &0.000 &0.000 \\ 
-0.085 &0.147 &-0.492 &0.000 &0.000 \\ 
0.000 &0.000 &0.000 &0.180 &0.055 \\ 
0.000 &0.000 &0.000 &0.055 &0.180 \\ 
\end{array}\right) $&$ \left( \begin{array}{ccccc}0.115 &-0.138 &0.000 &0.000 &0.000 \\ 
-0.138 &-0.044 &0.000 &0.000 &0.000 \\ 
0.000 &0.000 &0.478 &0.000 &0.000 \\ 
0.000 &0.000 &0.000 &0.455 &0.000 \\ 
0.000 &0.000 &0.000 &0.000 &0.478 \\ 
\end{array}\right) $&$ \left( \begin{array}{ccccc}0.026 &-0.006 &-0.013 &0.002 &-0.006 \\ 
-0.006 &0.019 &-0.014 &0.003 &-0.018 \\ 
-0.013 &-0.014 &0.016 &0.013 &-0.036 \\ 
0.002 &0.003 &0.013 &-0.003 &0.013 \\ 
-0.006 &-0.018 &-0.036 &0.013 &0.016 \\ 
\end{array}\right) $&$ \left( \begin{array}{ccccc}0.019 &0.011 &-0.023 &0.000 &0.000 \\ 
0.011 &0.006 &0.040 &0.000 &0.000 \\ 
-0.023 &0.040 &-0.087 &0.000 &0.000 \\ 
0.000 &0.000 &0.000 &0.025 &-0.001 \\ 
0.000 &0.000 &0.000 &-0.001 &0.025 \\ 
\end{array}\right) $&$ \left( \begin{array}{ccccc}0.002 &-0.016 &0.013 &0.000 &0.000 \\ 
-0.016 &-0.031 &0.047 &0.000 &0.000 \\ 
0.013 &0.047 &0.006 &0.000 &0.000 \\ 
0.000 &0.000 &0.000 &0.007 &0.000 \\ 
0.000 &0.000 &0.000 &0.000 &0.009 \\ 
\end{array}\right) $\\ 
\hline  
\end{tabular}  
}  
\end{sidewaystable*}  

%% file: EccsUS_summary_Latex_5
\begin{table*}[b]  
\caption{\label{tab:ECCsUS_summary_all} 
$\Xi$ matrix contributed from uniform starin for the four JT active centers in unconvention unit cell. JTC1, JTC2, JTC3 are the label for the JT active centers, as defined in Table. \ref{Table:coordinates_SAJTmodes_Oct}. From the left to right, and from the top to bottom, the matrix elements correspond to JTD\_2, JTD\_3, JTD\_4, JTD\_5 and JTD\_6, respectively. The unit is $bohr^2/Ry$.} 
\resizebox{\textwidth}{!}{  
\begin{tabular}{|c|c|c|}\hline  
& (0,0,0) (JTC1,JTC1) & (0,0,0) (JTC1,JTC2)  \\ \hline  
NaOs & $ \left( \begin{array}{ccccc}0.267 &0.000 &0.000 &0.000 &0.000 \\ 
0.000 &0.267 &0.000 &0.000 &0.000 \\ 
0.000 &0.000 &0.313 &0.000 &0.000 \\ 
0.000 &0.000 &0.000 &0.313 &0.000 \\ 
0.000 &0.000 &0.000 &0.000 &0.313 \\ 
\end{array}\right) $&$ \left( \begin{array}{ccccc}0.033 &-0.081 &0.000 &0.000 &0.000 \\ 
-0.081 &0.127 &0.000 &0.000 &0.000 \\ 
0.000 &0.000 &-0.016 &0.000 &0.000 \\ 
0.000 &0.000 &0.000 &0.148 &0.000 \\ 
0.000 &0.000 &0.000 &0.000 &0.148 \\ 
\end{array}\right) $\\ 
\hline  
LiOs & $ \left( \begin{array}{ccccc}0.386 &0.000 &0.000 &0.000 &0.000 \\ 
0.000 &0.386 &0.000 &0.000 &0.000 \\ 
0.000 &0.000 &0.243 &0.000 &0.000 \\ 
0.000 &0.000 &0.000 &0.243 &0.000 \\ 
0.000 &0.000 &0.000 &0.000 &0.243 \\ 
\end{array}\right) $&$ \left( \begin{array}{ccccc}0.047 &-0.117 &0.000 &0.000 &0.000 \\ 
-0.117 &0.182 &0.000 &0.000 &0.000 \\ 
0.000 &0.000 &-0.013 &0.000 &0.000 \\ 
0.000 &0.000 &0.000 &0.115 &0.000 \\ 
0.000 &0.000 &0.000 &0.000 &0.115 \\ 
\end{array}\right) $\\ 
\hline  
MgRe & $ \left( \begin{array}{ccccc}0.217 &0.000 &0.000 &0.000 &0.000 \\ 
0.000 &0.217 &0.000 &0.000 &0.000 \\ 
0.000 &0.000 &0.170 &0.000 &0.000 \\ 
0.000 &0.000 &0.000 &0.170 &0.000 \\ 
0.000 &0.000 &0.000 &0.000 &0.170 \\ 
\end{array}\right) $&$ \left( \begin{array}{ccccc}0.026 &-0.066 &0.000 &0.000 &0.000 \\ 
-0.066 &0.102 &0.000 &0.000 &0.000 \\ 
0.000 &0.000 &-0.009 &0.000 &0.000 \\ 
0.000 &0.000 &0.000 &0.080 &0.000 \\ 
0.000 &0.000 &0.000 &0.000 &0.080 \\ 
\end{array}\right) $\\ 
\hline  
ZnRe & $ \left( \begin{array}{ccccc}0.321 &0.000 &0.000 &0.000 &0.000 \\ 
0.000 &0.321 &0.000 &0.000 &0.000 \\ 
0.000 &0.000 &0.210 &0.000 &0.000 \\ 
0.000 &0.000 &0.000 &0.210 &0.000 \\ 
0.000 &0.000 &0.000 &0.000 &0.210 \\ 
\end{array}\right) $&$ \left( \begin{array}{ccccc}0.039 &-0.098 &0.000 &0.000 &0.000 \\ 
-0.098 &0.152 &0.000 &0.000 &0.000 \\ 
0.000 &0.000 &-0.012 &0.000 &0.000 \\ 
0.000 &0.000 &0.000 &0.099 &0.000 \\ 
0.000 &0.000 &0.000 &0.000 &0.099 \\ 
\end{array}\right) $\\ 
\hline  
TaCl & $ \left( \begin{array}{ccccc}4.819 &0.000 &0.000 &0.000 &0.000 \\ 
0.000 &4.819 &0.000 &0.000 &0.000 \\ 
0.000 &0.000 &3.136 &0.000 &0.000 \\ 
0.000 &0.000 &0.000 &3.136 &0.000 \\ 
0.000 &0.000 &0.000 &0.000 &3.136 \\ 
\end{array}\right) $&$ \left( \begin{array}{ccccc}0.628 &-1.452 &0.000 &0.000 &0.000 \\ 
-1.452 &2.305 &0.000 &0.000 &0.000 \\ 
0.000 &0.000 &-0.137 &0.000 &0.000 \\ 
0.000 &0.000 &0.000 &1.500 &0.000 \\ 
0.000 &0.000 &0.000 &0.000 &1.500 \\ 
\end{array}\right) $\\ 
\hline  
TaBr & $ \left( \begin{array}{ccccc}6.649 &0.000 &0.000 &0.000 &0.000 \\ 
0.000 &6.649 &0.000 &0.000 &0.000 \\ 
0.000 &0.000 &3.830 &0.000 &0.000 \\ 
0.000 &0.000 &0.000 &3.830 &0.000 \\ 
0.000 &0.000 &0.000 &0.000 &3.830 \\ 
\end{array}\right) $&$ \left( \begin{array}{ccccc}0.873 &-2.001 &0.000 &0.000 &0.000 \\ 
-2.001 &3.184 &0.000 &0.000 &0.000 \\ 
0.000 &0.000 &-0.162 &0.000 &0.000 \\ 
0.000 &0.000 &0.000 &1.834 &0.000 \\ 
0.000 &0.000 &0.000 &0.000 &1.834 \\ 
\end{array}\right) $\\ 
\hline  
YW & $ \left( \begin{array}{ccccc}0.084 &0.000 &0.000 &0.000 &0.000 \\ 
0.000 &0.084 &0.000 &0.000 &0.000 \\ 
0.000 &0.000 &0.138 &0.000 &0.000 \\ 
0.000 &0.000 &0.000 &0.138 &0.000 \\ 
0.000 &0.000 &0.000 &0.000 &0.138 \\ 
\end{array}\right) $&$ \left( \begin{array}{ccccc}0.010 &-0.026 &0.000 &0.000 &0.000 \\ 
-0.026 &0.040 &0.000 &0.000 &0.000 \\ 
0.000 &0.000 &-0.007 &0.000 &0.000 \\ 
0.000 &0.000 &0.000 &0.065 &0.000 \\ 
0.000 &0.000 &0.000 &0.000 &0.065 \\ 
\end{array}\right) $\\ 
\hline  
\end{tabular}  
}  
\end{table*}

%% file: B11Kappa_real_summary_Latex_8
\begin{sidewaystable*}[b]  
\caption{\label{tab:K_real_summary_allB11} 
Contribution from all phonons: $K ^{\mu \gamma}_{\mu' \gamma'} (\bm{R})$ in real space ($\mu,\mu'$ donate the four JT centers, $\gamma, \gamma' $ donate the two components, $\theta, \epsilon$ of $e_g$ JT active mode, and $\bm{R}$ represents the lattice vector). `000', `100', `110', `111' represent the original, nearest, next nearest convention unit cells. The unit is meV.} 
\resizebox{\textwidth}{!}{  
\begin{tabular}{|c|c|c|c|c|}\hline  
&  000 &  100 &  110 &  111  \\   
\hline  
NaOs & $ \left( \begin{array}{cccccccc}71.176 &0.000 &-0.903 &-0.487 &-0.899 &0.503 &-0.020 &-0.001 \\ 
0.000 &71.176 &-0.487 &-0.340 &0.489 &-0.325 &-0.001 &-1.187 \\ 
-0.903 &-0.487 &71.176 &0.000 &-0.058 &-0.001 &-0.899 &0.489 \\ 
-0.487 &-0.340 &0.000 &71.176 &-0.001 &-1.183 &0.503 &-0.325 \\ 
-0.899 &0.489 &-0.058 &-0.001 &71.176 &0.000 &-0.903 &-0.487 \\ 
0.503 &-0.325 &-0.001 &-1.183 &0.000 &71.176 &-0.487 &-0.340 \\ 
-0.020 &-0.001 &-0.899 &0.503 &-0.903 &-0.487 &71.176 &0.000 \\ 
-0.001 &-1.187 &0.489 &-0.325 &-0.487 &-0.340 &0.000 &71.176 \\ 
\end{array}\right) $&$ \left( \begin{array}{cccccccc}-0.338 &-0.866 &-0.099 &-0.064 &-0.902 &0.505 &-0.025 &0.003 \\ 
-0.866 &0.662 &-0.064 &-0.026 &0.492 &-0.334 &0.003 &-1.194 \\ 
-0.099 &-0.064 &-0.338 &-0.866 &-0.063 &0.003 &-0.902 &0.492 \\ 
-0.064 &-0.026 &-0.866 &0.662 &0.003 &-1.190 &0.505 &-0.334 \\ 
-0.902 &0.492 &-0.063 &0.003 &-0.338 &-0.866 &-0.099 &-0.064 \\ 
0.505 &-0.334 &0.003 &-1.190 &-0.866 &0.662 &-0.064 &-0.026 \\ 
-0.025 &0.003 &-0.902 &0.505 &-0.099 &-0.064 &-0.338 &-0.866 \\ 
0.003 &-1.194 &0.492 &-0.334 &-0.064 &-0.026 &-0.866 &0.662 \\ 
\end{array}\right) $&$ \left( \begin{array}{cccccccc}0.114 &0.000 &-0.102 &-0.045 &-0.103 &0.048 &-0.029 &-0.000 \\ 
0.000 &-0.241 &-0.063 &-0.036 &0.062 &-0.033 &-0.000 &-1.202 \\ 
-0.102 &-0.063 &0.115 &-0.005 &0.076 &0.140 &-0.103 &0.062 \\ 
-0.045 &-0.036 &-0.005 &-0.335 &0.140 &0.185 &0.048 &-0.033 \\ 
-0.103 &0.062 &0.076 &0.140 &0.115 &-0.005 &-0.102 &-0.063 \\ 
0.048 &-0.033 &0.140 &0.185 &-0.005 &-0.335 &-0.045 &-0.036 \\ 
-0.029 &-0.000 &-0.103 &0.048 &-0.102 &-0.045 &0.114 &0.000 \\ 
-0.000 &-1.202 &0.062 &-0.033 &-0.063 &-0.036 &0.000 &-0.241 \\ 
\end{array}\right) $&$ \left( \begin{array}{cccccccc}-0.025 &0.000 &-0.101 &-0.046 &-0.103 &0.049 &-0.016 &-0.000 \\ 
0.000 &-0.025 &-0.046 &-0.048 &0.045 &-0.045 &-0.000 &-0.131 \\ 
-0.101 &-0.046 &-0.040 &0.008 &-0.037 &-0.016 &-0.008 &-0.005 \\ 
-0.046 &-0.048 &0.008 &-0.050 &0.009 &0.003 &-0.024 &-0.017 \\ 
-0.103 &0.045 &-0.037 &0.009 &-0.040 &-0.009 &-0.007 &0.005 \\ 
0.049 &-0.045 &-0.016 &0.003 &-0.009 &-0.050 &0.028 &-0.018 \\ 
-0.016 &-0.000 &-0.008 &-0.024 &-0.007 &0.028 &-0.056 &-0.000 \\ 
-0.000 &-0.131 &-0.005 &-0.017 &0.005 &-0.018 &-0.000 &-0.034 \\ 
\end{array}\right) $\\ 
\hline  
LiOs & $ \left( \begin{array}{cccccccc}68.005 &0.000 &-0.901 &-0.759 &-0.906 &0.780 &0.431 &0.000 \\ 
0.000 &68.005 &-0.759 &-0.024 &0.754 &-0.019 &0.000 &-1.358 \\ 
-0.901 &-0.760 &68.015 &0.008 &0.394 &-0.016 &-0.907 &0.759 \\ 
-0.760 &-0.024 &0.008 &68.006 &0.015 &-1.334 &0.750 &-0.034 \\ 
-0.905 &0.755 &0.394 &-0.016 &68.015 &-0.008 &-0.902 &-0.764 \\ 
0.779 &-0.020 &0.014 &-1.334 &-0.008 &68.006 &-0.732 &-0.038 \\ 
0.429 &0.000 &-0.908 &0.784 &-0.902 &-0.763 &68.001 &0.000 \\ 
0.000 &-1.357 &0.729 &-0.033 &-0.733 &-0.038 &0.000 &68.019 \\ 
\end{array}\right) $&$ \left( \begin{array}{cccccccc}-0.388 &-0.384 &-0.126 &-0.074 &-0.907 &0.780 &0.430 &0.001 \\ 
-0.384 &0.055 &-0.074 &-0.041 &0.755 &-0.021 &0.001 &-1.360 \\ 
-0.127 &-0.075 &-0.385 &-0.395 &0.393 &-0.015 &-0.908 &0.760 \\ 
-0.075 &-0.040 &-0.395 &0.071 &0.016 &-1.336 &0.751 &-0.036 \\ 
-0.906 &0.756 &0.394 &-0.015 &-0.386 &-0.393 &-0.125 &-0.077 \\ 
0.779 &-0.022 &0.015 &-1.336 &-0.394 &0.059 &-0.063 &-0.043 \\ 
0.428 &0.001 &-0.909 &0.784 &-0.124 &-0.075 &-0.393 &-0.389 \\ 
0.001 &-1.358 &0.730 &-0.035 &-0.065 &-0.044 &-0.389 &0.066 \\ 
\end{array}\right) $&$ \left( \begin{array}{cccccccc}0.183 &-0.000 &-0.116 &-0.039 &-0.122 &0.053 &0.431 &0.000 \\ 
-0.000 &-0.046 &-0.084 &-0.040 &0.080 &-0.042 &0.000 &-1.365 \\ 
-0.116 &-0.086 &0.182 &-0.001 &0.011 &0.044 &-0.120 &0.083 \\ 
-0.040 &-0.040 &-0.001 &-0.155 &0.038 &0.084 &0.041 &-0.046 \\ 
-0.122 &0.082 &0.011 &0.044 &0.182 &-0.001 &-0.114 &-0.087 \\ 
0.053 &-0.043 &0.037 &0.084 &-0.001 &-0.155 &-0.028 &-0.044 \\ 
0.429 &0.000 &-0.120 &0.055 &-0.114 &-0.041 &0.186 &-0.000 \\ 
0.000 &-1.364 &0.071 &-0.046 &-0.075 &-0.044 &-0.000 &-0.038 \\ 
\end{array}\right) $&$ \left( \begin{array}{cccccccc}-0.009 &-0.000 &-0.103 &-0.053 &-0.109 &0.066 &-0.011 &0.000 \\ 
-0.000 &-0.009 &-0.053 &-0.043 &0.049 &-0.045 &0.000 &-0.152 \\ 
-0.103 &-0.053 &-0.020 &0.011 &-0.020 &-0.020 &-0.006 &0.021 \\ 
-0.052 &-0.043 &0.011 &-0.032 &0.028 &-0.008 &-0.023 &-0.018 \\ 
-0.109 &0.049 &-0.020 &0.030 &-0.020 &-0.011 &-0.006 &-0.021 \\ 
0.065 &-0.045 &-0.022 &-0.008 &-0.011 &-0.032 &0.025 &-0.018 \\ 
-0.011 &0.000 &-0.006 &-0.025 &-0.006 &0.027 &-0.040 &-0.000 \\ 
0.000 &-0.151 &0.023 &-0.018 &-0.023 &-0.018 &-0.000 &-0.013 \\ 
\end{array}\right) $\\ 
\hline  
MgRe & $ \left( \begin{array}{cccccccc}67.030 &0.001 &-1.803 &-1.066 &-1.800 &1.123 &0.157 &-0.003 \\ 
0.001 &67.029 &-1.066 &-0.572 &1.067 &-0.527 &-0.003 &-2.444 \\ 
-1.803 &-1.066 &67.030 &0.001 &0.050 &-0.005 &-1.800 &1.067 \\ 
-1.066 &-0.572 &0.001 &67.030 &-0.005 &-2.416 &1.123 &-0.527 \\ 
-1.800 &1.067 &0.050 &-0.005 &67.030 &0.001 &-1.803 &-1.066 \\ 
1.123 &-0.527 &-0.005 &-2.416 &0.001 &67.030 &-1.066 &-0.572 \\ 
0.157 &-0.003 &-1.800 &1.123 &-1.803 &-1.066 &67.030 &0.001 \\ 
-0.003 &-2.444 &1.067 &-0.527 &-1.066 &-0.572 &0.001 &67.029 \\ 
\end{array}\right) $&$ \left( \begin{array}{cccccccc}-0.097 &-0.586 &-0.208 &-0.214 &-1.805 &1.130 &0.148 &0.006 \\ 
-0.586 &0.579 &-0.214 &0.040 &1.074 &-0.544 &0.006 &-2.458 \\ 
-0.208 &-0.214 &-0.097 &-0.587 &0.042 &0.004 &-1.805 &1.074 \\ 
-0.214 &0.040 &-0.587 &0.581 &0.004 &-2.430 &1.130 &-0.544 \\ 
-1.805 &1.074 &0.042 &0.004 &-0.097 &-0.587 &-0.208 &-0.214 \\ 
1.130 &-0.544 &0.004 &-2.430 &-0.587 &0.581 &-0.214 &0.040 \\ 
0.148 &0.006 &-1.805 &1.130 &-0.208 &-0.214 &-0.097 &-0.586 \\ 
0.006 &-2.458 &1.074 &-0.544 &-0.214 &0.040 &-0.586 &0.579 \\ 
\end{array}\right) $&$ \left( \begin{array}{cccccccc}0.331 &0.000 &-0.204 &-0.122 &-0.211 &0.149 &0.149 &-0.002 \\ 
0.000 &0.113 &-0.228 &-0.009 &0.225 &0.013 &-0.002 &-2.489 \\ 
-0.204 &-0.228 &0.341 &-0.008 &0.071 &0.114 &-0.211 &0.225 \\ 
-0.122 &-0.009 &-0.008 &-0.285 &0.114 &0.195 &0.149 &0.013 \\ 
-0.211 &0.225 &0.071 &0.114 &0.341 &-0.008 &-0.204 &-0.228 \\ 
0.149 &0.013 &0.114 &0.195 &-0.008 &-0.285 &-0.122 &-0.009 \\ 
0.149 &-0.002 &-0.211 &0.149 &-0.204 &-0.122 &0.331 &0.000 \\ 
-0.002 &-2.489 &0.225 &0.013 &-0.228 &-0.009 &0.000 &0.113 \\ 
\end{array}\right) $&$ \left( \begin{array}{cccccccc}0.076 &0.000 &-0.205 &-0.138 &-0.212 &0.165 &0.076 &-0.001 \\ 
0.000 &0.075 &-0.139 &-0.046 &0.135 &-0.026 &-0.001 &-0.307 \\ 
-0.205 &-0.139 &0.010 &0.039 &-0.016 &-0.049 &0.012 &0.057 \\ 
-0.138 &-0.046 &0.039 &-0.035 &0.072 &0.006 &-0.049 &-0.007 \\ 
-0.212 &0.135 &-0.016 &0.072 &0.011 &-0.042 &0.015 &-0.058 \\ 
0.165 &-0.026 &-0.049 &0.006 &-0.042 &-0.036 &0.054 &-0.012 \\ 
0.076 &-0.001 &0.012 &-0.049 &0.015 &0.054 &-0.063 &0.000 \\ 
-0.001 &-0.307 &0.057 &-0.007 &-0.058 &-0.012 &0.000 &0.037 \\ 
\end{array}\right) $\\ 
\hline  
ZnRe & $ \left( \begin{array}{cccccccc}64.945 &0.001 &-1.630 &-1.318 &-1.649 &1.386 &0.706 &0.001 \\ 
0.001 &64.943 &-1.319 &-0.107 &1.302 &-0.092 &0.001 &-2.456 \\ 
-1.630 &-1.319 &64.945 &0.001 &0.659 &-0.004 &-1.649 &1.302 \\ 
-1.318 &-0.107 &0.001 &64.945 &-0.004 &-2.391 &1.386 &-0.092 \\ 
-1.649 &1.302 &0.659 &-0.004 &64.945 &0.001 &-1.630 &-1.319 \\ 
1.386 &-0.092 &-0.004 &-2.391 &0.001 &64.945 &-1.318 &-0.107 \\ 
0.706 &0.001 &-1.649 &1.386 &-1.630 &-1.318 &64.945 &0.001 \\ 
0.001 &-2.456 &1.302 &-0.092 &-1.319 &-0.107 &0.001 &64.943 \\ 
\end{array}\right) $&$ \left( \begin{array}{cccccccc}0.016 &-0.184 &-0.367 &-0.280 &-1.655 &1.392 &0.703 &0.004 \\ 
-0.184 &0.228 &-0.280 &-0.044 &1.307 &-0.099 &0.004 &-2.467 \\ 
-0.367 &-0.280 &0.016 &-0.184 &0.658 &-0.001 &-1.655 &1.307 \\ 
-0.280 &-0.044 &-0.184 &0.230 &-0.001 &-2.402 &1.392 &-0.099 \\ 
-1.655 &1.307 &0.658 &-0.001 &0.016 &-0.184 &-0.367 &-0.280 \\ 
1.392 &-0.099 &-0.001 &-2.402 &-0.184 &0.230 &-0.280 &-0.044 \\ 
0.703 &0.004 &-1.655 &1.392 &-0.367 &-0.280 &0.016 &-0.184 \\ 
0.004 &-2.467 &1.307 &-0.099 &-0.280 &-0.044 &-0.184 &0.228 \\ 
\end{array}\right) $&$ \left( \begin{array}{cccccccc}0.515 &-0.000 &-0.334 &-0.156 &-0.350 &0.200 &0.712 &0.002 \\ 
-0.000 &0.181 &-0.324 &-0.053 &0.313 &-0.054 &0.002 &-2.502 \\ 
-0.334 &-0.324 &0.511 &-0.002 &0.004 &0.015 &-0.350 &0.313 \\ 
-0.156 &-0.053 &-0.002 &-0.241 &0.015 &0.147 &0.200 &-0.054 \\ 
-0.350 &0.313 &0.004 &0.015 &0.511 &-0.002 &-0.334 &-0.324 \\ 
0.200 &-0.054 &0.015 &0.147 &-0.002 &-0.241 &-0.156 &-0.053 \\ 
0.712 &0.002 &-0.350 &0.200 &-0.334 &-0.156 &0.515 &-0.000 \\ 
0.002 &-2.502 &0.313 &-0.054 &-0.324 &-0.053 &-0.000 &0.181 \\ 
\end{array}\right) $&$ \left( \begin{array}{cccccccc}0.145 &-0.000 &-0.303 &-0.210 &-0.320 &0.252 &0.074 &0.001 \\ 
-0.000 &0.146 &-0.210 &-0.062 &0.197 &-0.062 &0.001 &-0.475 \\ 
-0.303 &-0.210 &0.045 &0.057 &-0.027 &-0.072 &-0.008 &0.101 \\ 
-0.210 &-0.062 &0.057 &-0.021 &0.114 &-0.027 &-0.069 &-0.044 \\ 
-0.320 &0.197 &-0.027 &0.114 &0.045 &-0.060 &-0.006 &-0.101 \\ 
0.252 &-0.062 &-0.072 &-0.027 &-0.060 &-0.022 &0.075 &-0.040 \\ 
0.074 &0.001 &-0.008 &-0.069 &-0.006 &0.075 &-0.061 &-0.000 \\ 
0.001 &-0.475 &0.101 &-0.044 &-0.101 &-0.040 &-0.000 &0.083 \\ 
\end{array}\right) $\\ 
\hline  
TaCl & $ \left( \begin{array}{cccccccc}32.784 &0.000 &-0.245 &-0.147 &-0.245 &0.150 &0.014 &-0.000 \\ 
0.000 &32.784 &-0.147 &-0.075 &0.147 &-0.074 &-0.000 &-0.332 \\ 
-0.245 &-0.147 &32.784 &0.000 &0.010 &-0.000 &-0.245 &0.147 \\ 
-0.147 &-0.075 &0.000 &32.784 &-0.000 &-0.330 &0.150 &-0.074 \\ 
-0.245 &0.147 &0.010 &-0.000 &32.784 &0.000 &-0.245 &-0.147 \\ 
0.150 &-0.074 &-0.000 &-0.330 &0.000 &32.784 &-0.147 &-0.075 \\ 
0.014 &-0.000 &-0.245 &0.150 &-0.245 &-0.147 &32.784 &0.000 \\ 
-0.000 &-0.332 &0.147 &-0.074 &-0.147 &-0.075 &0.000 &32.784 \\ 
\end{array}\right) $&$ \left( \begin{array}{cccccccc}-0.315 &-0.077 &-0.022 &-0.025 &-0.245 &0.150 &0.014 &-0.000 \\ 
-0.077 &-0.226 &-0.025 &0.006 &0.147 &-0.074 &-0.000 &-0.332 \\ 
-0.022 &-0.025 &-0.315 &-0.077 &0.010 &-0.000 &-0.245 &0.147 \\ 
-0.025 &0.006 &-0.077 &-0.226 &-0.000 &-0.330 &0.150 &-0.074 \\ 
-0.245 &0.147 &0.010 &-0.000 &-0.315 &-0.077 &-0.022 &-0.025 \\ 
0.150 &-0.074 &-0.000 &-0.330 &-0.077 &-0.226 &-0.025 &0.006 \\ 
0.014 &-0.000 &-0.245 &0.150 &-0.022 &-0.025 &-0.315 &-0.077 \\ 
-0.000 &-0.332 &0.147 &-0.074 &-0.025 &0.006 &-0.077 &-0.226 \\ 
\end{array}\right) $&$ \left( \begin{array}{cccccccc}0.051 &0.000 &-0.022 &-0.021 &-0.022 &0.022 &0.014 &-0.000 \\ 
0.000 &-0.015 &-0.025 &0.005 &0.025 &0.005 &-0.000 &-0.332 \\ 
-0.022 &-0.025 &0.051 &-0.000 &0.010 &0.018 &-0.022 &0.025 \\ 
-0.021 &0.005 &-0.000 &-0.030 &0.018 &0.031 &0.022 &0.005 \\ 
-0.022 &0.025 &0.010 &0.018 &0.051 &-0.000 &-0.022 &-0.025 \\ 
0.022 &0.005 &0.018 &0.031 &-0.000 &-0.030 &-0.021 &0.005 \\ 
0.014 &-0.000 &-0.022 &0.022 &-0.022 &-0.021 &0.051 &0.000 \\ 
-0.000 &-0.332 &0.025 &0.005 &-0.025 &0.005 &0.000 &-0.015 \\ 
\end{array}\right) $&$ \left( \begin{array}{cccccccc}-0.013 &0.000 &-0.022 &-0.021 &-0.022 &0.022 &0.017 &-0.000 \\ 
0.000 &-0.013 &-0.021 &0.003 &0.021 &0.004 &-0.000 &-0.035 \\ 
-0.022 &-0.021 &-0.016 &0.001 &-0.011 &-0.001 &0.003 &-0.005 \\ 
-0.021 &0.003 &0.001 &-0.017 &0.004 &0.006 &-0.009 &-0.007 \\ 
-0.022 &0.021 &-0.011 &0.004 &-0.016 &-0.001 &0.003 &0.004 \\ 
0.022 &0.004 &-0.001 &0.006 &-0.001 &-0.017 &0.009 &-0.007 \\ 
0.017 &-0.000 &0.003 &-0.009 &0.003 &0.009 &-0.018 &-0.000 \\ 
-0.000 &-0.035 &-0.005 &-0.007 &0.004 &-0.007 &-0.000 &-0.015 \\ 
\end{array}\right) $\\ 
\hline  
TaBr & $ \left( \begin{array}{cccccccc}29.949 &0.000 &-0.264 &-0.166 &-0.264 &0.169 &0.028 &-0.000 \\ 
0.000 &29.949 &-0.166 &-0.073 &0.166 &-0.071 &-0.000 &-0.362 \\ 
-0.264 &-0.166 &29.949 &0.000 &0.023 &-0.000 &-0.264 &0.166 \\ 
-0.166 &-0.073 &0.000 &29.949 &-0.000 &-0.360 &0.169 &-0.071 \\ 
-0.264 &0.166 &0.023 &-0.000 &29.949 &0.000 &-0.264 &-0.166 \\ 
0.169 &-0.071 &-0.000 &-0.360 &0.000 &29.949 &-0.166 &-0.073 \\ 
0.028 &-0.000 &-0.264 &0.169 &-0.264 &-0.166 &29.949 &0.000 \\ 
-0.000 &-0.362 &0.166 &-0.071 &-0.166 &-0.073 &0.000 &29.949 \\ 
\end{array}\right) $&$ \left( \begin{array}{cccccccc}-0.290 &-0.087 &-0.022 &-0.024 &-0.265 &0.169 &0.028 &-0.000 \\ 
-0.087 &-0.189 &-0.024 &0.005 &0.165 &-0.071 &-0.000 &-0.362 \\ 
-0.022 &-0.024 &-0.290 &-0.087 &0.023 &-0.000 &-0.265 &0.165 \\ 
-0.024 &0.005 &-0.087 &-0.189 &-0.000 &-0.360 &0.169 &-0.071 \\ 
-0.265 &0.165 &0.023 &-0.000 &-0.290 &-0.087 &-0.022 &-0.024 \\ 
0.169 &-0.071 &-0.000 &-0.360 &-0.087 &-0.189 &-0.024 &0.005 \\ 
0.028 &-0.000 &-0.265 &0.169 &-0.022 &-0.024 &-0.290 &-0.087 \\ 
-0.000 &-0.362 &0.165 &-0.071 &-0.024 &0.005 &-0.087 &-0.189 \\ 
\end{array}\right) $&$ \left( \begin{array}{cccccccc}0.045 &0.000 &-0.021 &-0.019 &-0.022 &0.020 &0.028 &-0.000 \\ 
0.000 &-0.015 &-0.024 &0.003 &0.024 &0.004 &-0.000 &-0.362 \\ 
-0.021 &-0.024 &0.045 &-0.000 &0.007 &0.018 &-0.022 &0.024 \\ 
-0.019 &0.003 &-0.000 &-0.035 &0.018 &0.033 &0.020 &0.004 \\ 
-0.022 &0.024 &0.007 &0.018 &0.045 &-0.000 &-0.021 &-0.024 \\ 
0.020 &0.004 &0.018 &0.033 &-0.000 &-0.035 &-0.019 &0.003 \\ 
0.028 &-0.000 &-0.022 &0.020 &-0.021 &-0.019 &0.045 &0.000 \\ 
-0.000 &-0.362 &0.024 &0.004 &-0.024 &0.003 &0.000 &-0.015 \\ 
\end{array}\right) $&$ \left( \begin{array}{cccccccc}-0.008 &0.000 &-0.021 &-0.019 &-0.021 &0.021 &0.014 &-0.000 \\ 
0.000 &-0.008 &-0.019 &0.002 &0.019 &0.002 &-0.000 &-0.033 \\ 
-0.021 &-0.019 &-0.011 &0.002 &-0.009 &-0.002 &0.003 &-0.002 \\ 
-0.019 &0.002 &0.002 &-0.013 &0.005 &0.005 &-0.008 &-0.006 \\ 
-0.021 &0.019 &-0.009 &0.005 &-0.011 &-0.002 &0.003 &0.002 \\ 
0.021 &0.002 &-0.002 &0.005 &-0.002 &-0.013 &0.008 &-0.006 \\ 
0.014 &-0.000 &0.003 &-0.008 &0.003 &0.008 &-0.014 &-0.000 \\ 
-0.000 &-0.033 &-0.002 &-0.006 &0.002 &-0.006 &-0.000 &-0.010 \\ 
\end{array}\right) $\\ 
\hline  
YW & $ \left( \begin{array}{cccccccc}69.120 &0.001 &-1.948 &-0.834 &-1.918 &0.881 &-0.321 &-0.005 \\ 
0.001 &69.118 &-0.835 &-0.984 &0.850 &-0.904 &-0.005 &-2.403 \\ 
-1.948 &-0.835 &69.120 &0.001 &-0.494 &-0.006 &-1.918 &0.850 \\ 
-0.834 &-0.984 &0.001 &69.119 &-0.006 &-2.423 &0.881 &-0.904 \\ 
-1.918 &0.850 &-0.494 &-0.006 &69.120 &0.001 &-1.948 &-0.835 \\ 
0.881 &-0.904 &-0.006 &-2.423 &0.001 &69.119 &-0.834 &-0.984 \\ 
-0.321 &-0.005 &-1.918 &0.881 &-1.948 &-0.834 &69.120 &0.001 \\ 
-0.005 &-2.403 &0.850 &-0.904 &-0.835 &-0.984 &0.001 &69.118 \\ 
\end{array}\right) $&$ \left( \begin{array}{cccccccc}0.018 &-1.847 &-0.170 &-0.101 &-1.931 &0.898 &-0.346 &0.018 \\ 
-1.847 &2.150 &-0.102 &-0.054 &0.867 &-0.953 &0.018 &-2.440 \\ 
-0.170 &-0.102 &0.019 &-1.848 &-0.519 &0.019 &-1.931 &0.867 \\ 
-0.101 &-0.054 &-1.848 &2.151 &0.019 &-2.461 &0.898 &-0.953 \\ 
-1.931 &0.867 &-0.519 &0.019 &0.019 &-1.848 &-0.170 &-0.102 \\ 
0.898 &-0.953 &0.019 &-2.461 &-1.848 &2.151 &-0.101 &-0.054 \\ 
-0.346 &0.018 &-1.931 &0.898 &-0.170 &-0.101 &0.018 &-1.847 \\ 
0.018 &-2.440 &0.867 &-0.953 &-0.102 &-0.054 &-1.847 &2.150 \\ 
\end{array}\right) $&$ \left( \begin{array}{cccccccc}-0.012 &0.001 &-0.212 &-0.033 &-0.200 &0.047 &-0.360 &-0.004 \\ 
0.001 &0.024 &-0.089 &-0.129 &0.090 &-0.093 &-0.004 &-2.486 \\ 
-0.212 &-0.089 &-0.026 &-0.019 &0.178 &0.201 &-0.200 &0.090 \\ 
-0.033 &-0.129 &-0.019 &-0.357 &0.201 &0.257 &0.047 &-0.093 \\ 
-0.200 &0.090 &0.178 &0.201 &-0.026 &-0.019 &-0.212 &-0.089 \\ 
0.047 &-0.093 &0.201 &0.257 &-0.019 &-0.357 &-0.033 &-0.129 \\ 
-0.360 &-0.004 &-0.200 &0.047 &-0.212 &-0.033 &-0.012 &0.001 \\ 
-0.004 &-2.486 &0.090 &-0.093 &-0.089 &-0.129 &0.001 &0.024 \\ 
\end{array}\right) $&$ \left( \begin{array}{cccccccc}0.023 &0.001 &-0.240 &-0.031 &-0.227 &0.046 &-0.124 &-0.002 \\ 
0.001 &0.022 &-0.032 &-0.206 &0.033 &-0.175 &-0.002 &-0.241 \\ 
-0.240 &-0.032 &-0.047 &0.027 &0.015 &-0.047 &0.005 &0.040 \\ 
-0.031 &-0.206 &0.027 &-0.078 &0.020 &0.006 &-0.014 &0.049 \\ 
-0.227 &0.033 &0.015 &0.020 &-0.046 &-0.030 &0.005 &-0.038 \\ 
0.046 &-0.175 &-0.047 &0.006 &-0.030 &-0.080 &0.025 &0.040 \\ 
-0.124 &-0.002 &0.005 &-0.014 &0.005 &0.025 &-0.101 &0.000 \\ 
-0.002 &-0.241 &0.040 &0.049 &-0.038 &0.040 &0.000 &-0.026 \\ 
\end{array}\right) $\\ 
\hline  
\end{tabular}  
}  
\end{sidewaystable*}  

%% file: USKappa_real_summary_Latex_8
\begin{table*}[b]  
\caption{\label{tab:K_real_summary_allUS} 
Contribution from uniform strain: $K ^{\mu \gamma}_{\mu' \gamma'} $ matrix in real space. The each 2$\times$2 block correspondd to $ \left( \begin{array}{cc} K ^{E_{g\theta}}_{E_{g\theta}} &K ^{E_{g\theta}}_{E_{g\epsilon}} \\ K ^{E_{g\epsilon}}_{E_{g\theta}} & K ^{E_{g\epsilon}}_{E_{g\epsilon}}  \\ \end{array}\right) $ of one JT center. The unit is meV.} 
\resizebox{0.5\textwidth}{!}{  
\begin{tabular}{|c|c|}\hline  
&  onsite  \\   
\hline  
NaOs & $ \left( \begin{array}{cccccccc}-14.040 &0.000 &-0.438 &4.712 &-0.438 &-4.712 &-8.599 &-0.000 \\ 
0.000 &-14.040 &4.712 &-5.878 &-4.712 &-5.878 &-0.000 &2.283 \\ 
-0.438 &4.712 &-2.061 &2.927 &1.102 &1.019 &0.664 &-0.766 \\ 
4.712 &-5.878 &2.927 &-5.441 &-1.019 &0.518 &1.271 &0.956 \\ 
-0.438 &-4.712 &1.102 &-1.019 &-2.061 &-2.927 &0.664 &0.766 \\ 
-4.712 &-5.878 &1.019 &0.518 &-2.927 &-5.441 &-1.271 &0.956 \\ 
-8.599 &-0.000 &0.664 &1.271 &0.664 &-1.271 &-7.131 &0.000 \\ 
-0.000 &2.283 &-0.766 &0.956 &0.766 &0.956 &0.000 &-0.371 \\ 
\end{array}\right) $\\ 
\hline  
LiOs & $ \left( \begin{array}{cccccccc}-15.532 &0.000 &-0.296 &5.283 &-0.283 &-5.283 &-9.481 &-0.001 \\ 
-0.000 &-15.532 &5.285 &-6.399 &-5.302 &-6.414 &0.002 &2.781 \\ 
-0.285 &5.309 &-2.184 &2.928 &1.437 &1.452 &0.567 &-0.951 \\ 
5.311 &-6.415 &2.928 &-5.565 &-1.452 &0.272 &1.959 &1.149 \\ 
-0.298 &-5.292 &1.421 &-1.432 &-2.184 &-2.933 &0.563 &0.947 \\ 
-5.311 &-6.399 &1.431 &0.287 &-2.923 &-5.565 &-1.957 &1.146 \\ 
-9.449 &-0.002 &0.567 &1.920 &0.571 &-1.923 &-7.255 &0.000 \\ 
0.001 &2.754 &-0.937 &1.134 &0.940 &1.138 &-0.000 &-0.493 \\ 
\end{array}\right) $\\ 
\hline  
MgRe & $ \left( \begin{array}{cccccccc}-12.159 &-0.000 &0.924 &4.532 &0.924 &-4.532 &-6.925 &0.000 \\ 
-0.000 &-12.159 &4.532 &-4.309 &-4.532 &-4.309 &-0.000 &3.541 \\ 
0.924 &4.532 &-2.168 &1.969 &1.211 &1.244 &1.343 &-1.320 \\ 
4.532 &-4.309 &1.969 &-4.441 &-1.244 &1.387 &1.168 &1.255 \\ 
0.924 &-4.532 &1.211 &-1.244 &-2.168 &-1.969 &1.343 &1.320 \\ 
-4.532 &-4.309 &1.244 &1.387 &-1.969 &-4.441 &-1.168 &1.255 \\ 
-6.925 &-0.000 &1.343 &1.168 &1.343 &-1.168 &-5.577 &0.000 \\ 
0.000 &3.541 &-1.320 &1.255 &1.320 &1.255 &0.000 &-1.031 \\ 
\end{array}\right) $\\ 
\hline  
ZnRe & $ \left( \begin{array}{cccccccc}-14.408 &0.000 &1.259 &5.427 &1.259 &-5.427 &-8.141 &-0.000 \\ 
0.000 &-14.408 &5.427 &-5.008 &-5.427 &-5.008 &0.000 &4.392 \\ 
1.259 &5.427 &-2.518 &2.043 &1.570 &1.729 &1.440 &-1.654 \\ 
5.427 &-5.008 &2.043 &-4.878 &-1.729 &1.396 &1.805 &1.527 \\ 
1.259 &-5.427 &1.570 &-1.729 &-2.518 &-2.043 &1.440 &1.654 \\ 
-5.427 &-5.008 &1.729 &1.396 &-2.043 &-4.878 &-1.805 &1.527 \\ 
-8.141 &0.000 &1.440 &1.805 &1.440 &-1.805 &-6.057 &-0.000 \\ 
-0.000 &4.392 &-1.654 &1.527 &1.654 &1.527 &-0.000 &-1.339 \\ 
\end{array}\right) $\\ 
\hline  
TaCl & $ \left( \begin{array}{cccccccc}-14.749 &0.000 &-1.118 &4.722 &-1.118 &-4.722 &-9.296 &-0.000 \\ 
0.000 &-14.749 &4.722 &-6.570 &-4.722 &-6.570 &-0.000 &1.609 \\ 
-1.118 &4.722 &-2.004 &3.168 &1.019 &1.039 &0.111 &-0.515 \\ 
4.722 &-6.570 &3.168 &-5.662 &-1.039 &-0.192 &1.564 &0.717 \\ 
-1.118 &-4.722 &1.019 &-1.039 &-2.004 &-3.168 &0.111 &0.515 \\ 
-4.722 &-6.570 &1.039 &-0.192 &-3.168 &-5.662 &-1.564 &0.717 \\ 
-9.296 &-0.000 &0.111 &1.564 &0.111 &-1.564 &-7.491 &-0.000 \\ 
-0.000 &1.609 &-0.515 &0.717 &0.515 &0.717 &-0.000 &-0.175 \\ 
\end{array}\right) $\\ 
\hline  
TaBr & $ \left( \begin{array}{cccccccc}-14.091 &-0.000 &-0.955 &4.550 &-0.955 &-4.550 &-8.837 &0.000 \\ 
-0.000 &-14.091 &4.550 &-6.210 &-4.550 &-6.210 &-0.000 &1.672 \\ 
-0.955 &4.550 &-1.922 &2.986 &1.017 &1.025 &0.177 &-0.540 \\ 
4.550 &-6.210 &2.986 &-5.370 &-1.025 &-0.103 &1.509 &0.737 \\ 
-0.955 &-4.550 &1.017 &-1.025 &-1.922 &-2.986 &0.177 &0.540 \\ 
-4.550 &-6.210 &1.025 &-0.103 &-2.986 &-5.370 &-1.509 &0.737 \\ 
-8.837 &-0.000 &0.177 &1.509 &0.177 &-1.509 &-7.094 &0.000 \\ 
0.000 &1.672 &-0.540 &0.737 &0.540 &0.737 &0.000 &-0.198 \\ 
\end{array}\right) $\\ 
\hline  
YW & $ \left( \begin{array}{cccccccc}-8.105 &-0.000 &0.680 &3.043 &0.680 &-3.043 &-4.591 &-0.000 \\ 
-0.000 &-8.105 &3.043 &-2.834 &-3.043 &-2.834 &0.000 &2.437 \\ 
0.680 &3.043 &-1.575 &1.459 &0.710 &0.669 &1.136 &-0.915 \\ 
3.043 &-2.834 &1.459 &-3.260 &-0.669 &1.278 &0.423 &0.852 \\ 
0.680 &-3.043 &0.710 &-0.669 &-1.575 &-1.459 &1.136 &0.915 \\ 
-3.043 &-2.834 &0.669 &1.278 &-1.459 &-3.260 &-0.423 &0.852 \\ 
-4.591 &0.000 &1.136 &0.423 &1.136 &-0.423 &-4.102 &0.000 \\ 
-0.000 &2.437 &-0.915 &0.852 &0.915 &0.852 &0.000 &-0.733 \\ 
\end{array}\right) $\\ 
\hline  
\end{tabular}  
}  
\end{table*}

%% file: IFCKappaKappa_real_summary_Latex_8
\begin{table*}[b]  
\caption{\label{tab:K_real_summary_allIFCKappa} 
K matix from IFCs:: $K ^{\mu \gamma}_{\mu' \gamma'} $ matrix in real space. The each 2$\times$2 block correspondd to $ \left( \begin{array}{cc} K ^{E_{g\theta}}_{E_{g\theta}} &K ^{E_{g\theta}}_{E_{g\epsilon}} \\ K ^{E_{g\epsilon}}_{E_{g\theta}} & K ^{E_{g\epsilon}}_{E_{g\epsilon}}  \\ \end{array}\right) $ of one JT center. The unit is $meV$.} 
\resizebox{0.5\textwidth}{!}{  
\begin{tabular}{|c|c|}\hline  
&  onsite  \\   
\hline  
NaOs & $ \left( \begin{array}{cccccccc}75.840 &0.000 &-1.059 &-1.040 &-1.059 &1.040 &0.743 &-0.000 \\ 
0.000 &75.840 &-1.040 &0.143 &1.040 &0.143 &-0.000 &-1.659 \\ 
-1.059 &-1.040 &75.840 &0.000 &0.743 &0.000 &-1.059 &1.040 \\ 
-1.040 &0.143 &0.000 &75.840 &-0.000 &-1.659 &1.040 &0.143 \\ 
-1.059 &1.040 &0.743 &-0.000 &75.840 &0.000 &-1.059 &-1.040 \\ 
1.040 &0.143 &-0.000 &-1.659 &0.000 &75.840 &-1.040 &0.143 \\ 
0.743 &0.000 &-1.059 &1.040 &-1.059 &-1.040 &75.840 &0.000 \\ 
-0.000 &-1.659 &1.040 &0.143 &-1.040 &0.143 &0.000 &75.840 \\ 
\end{array}\right) $\\ 
\hline  
LiOs & $ \left( \begin{array}{cccccccc}71.103 &0.000 &-1.366 &-1.249 &-1.366 &1.249 &0.799 &-0.000 \\ 
-0.000 &71.103 &-1.249 &0.077 &1.249 &0.077 &0.000 &-2.087 \\ 
-1.368 &-1.258 &71.112 &0.003 &0.798 &-0.002 &-1.366 &1.259 \\ 
-1.258 &0.084 &0.003 &71.108 &0.010 &-2.091 &1.247 &0.072 \\ 
-1.368 &1.258 &0.798 &0.002 &71.112 &-0.003 &-1.366 &-1.259 \\ 
1.258 &0.084 &-0.010 &-2.091 &-0.003 &71.108 &-1.247 &0.072 \\ 
0.810 &0.000 &-1.373 &1.243 &-1.373 &-1.243 &71.106 &0.000 \\ 
-0.000 &-2.095 &1.255 &0.079 &-1.255 &0.079 &0.000 &71.114 \\ 
\end{array}\right) $\\ 
\hline  
MgRe & $ \left( \begin{array}{cccccccc}71.577 &0.000 &-2.285 &-1.809 &-2.285 &1.809 &0.849 &-0.000 \\ 
0.000 &71.577 &-1.809 &-0.196 &1.809 &-0.196 &-0.000 &-3.329 \\ 
-2.285 &-1.809 &71.577 &0.000 &0.849 &-0.000 &-2.285 &1.809 \\ 
-1.809 &-0.196 &0.000 &71.577 &0.000 &-3.329 &1.809 &-0.196 \\ 
-2.285 &1.809 &0.849 &0.000 &71.577 &0.000 &-2.285 &-1.809 \\ 
1.809 &-0.196 &0.000 &-3.329 &0.000 &71.577 &-1.809 &-0.196 \\ 
0.849 &0.000 &-2.285 &1.809 &-2.285 &-1.809 &71.577 &0.000 \\ 
0.000 &-3.329 &1.809 &-0.196 &-1.809 &-0.196 &0.000 &71.577 \\ 
\end{array}\right) $\\ 
\hline  
ZnRe & $ \left( \begin{array}{cccccccc}69.865 &0.000 &-2.639 &-2.244 &-2.639 &2.244 &1.248 &0.000 \\ 
0.000 &69.865 &-2.244 &-0.047 &2.244 &-0.047 &-0.000 &-3.934 \\ 
-2.639 &-2.244 &69.865 &0.000 &1.248 &0.000 &-2.639 &2.244 \\ 
-2.244 &-0.047 &0.000 &69.865 &-0.000 &-3.934 &2.244 &-0.047 \\ 
-2.639 &2.244 &1.248 &0.000 &69.865 &0.000 &-2.639 &-2.244 \\ 
2.244 &-0.047 &-0.000 &-3.934 &0.000 &69.865 &-2.244 &-0.047 \\ 
1.248 &-0.000 &-2.639 &2.244 &-2.639 &-2.244 &69.865 &0.000 \\ 
-0.000 &-3.934 &2.244 &-0.047 &-2.244 &-0.047 &0.000 &69.865 \\ 
\end{array}\right) $\\ 
\hline  
TaCl & $ \left( \begin{array}{cccccccc}33.319 &0.000 &-0.297 &-0.247 &-0.297 &0.247 &0.130 &-0.000 \\ 
0.000 &33.319 &-0.247 &-0.012 &0.247 &-0.012 &-0.000 &-0.439 \\ 
-0.297 &-0.247 &33.319 &0.000 &0.130 &-0.000 &-0.297 &0.247 \\ 
-0.247 &-0.012 &0.000 &33.319 &-0.000 &-0.439 &0.247 &-0.012 \\ 
-0.297 &0.247 &0.130 &-0.000 &33.319 &0.000 &-0.297 &-0.247 \\ 
0.247 &-0.012 &-0.000 &-0.439 &0.000 &33.319 &-0.247 &-0.012 \\ 
0.130 &0.000 &-0.297 &0.247 &-0.297 &-0.247 &33.319 &0.000 \\ 
-0.000 &-0.439 &0.247 &-0.012 &-0.247 &-0.012 &0.000 &33.319 \\ 
\end{array}\right) $\\ 
\hline  
TaBr & $ \left( \begin{array}{cccccccc}30.482 &0.000 &-0.306 &-0.255 &-0.306 &0.255 &0.136 &0.000 \\ 
-0.000 &30.482 &-0.255 &-0.011 &0.255 &-0.011 &0.000 &-0.453 \\ 
-0.306 &-0.255 &30.482 &0.000 &0.136 &-0.000 &-0.306 &0.255 \\ 
-0.255 &-0.011 &-0.000 &30.482 &-0.000 &-0.453 &0.255 &-0.011 \\ 
-0.306 &0.255 &0.136 &-0.000 &30.482 &0.000 &-0.306 &-0.255 \\ 
0.255 &-0.011 &-0.000 &-0.453 &-0.000 &30.482 &-0.255 &-0.011 \\ 
0.136 &0.000 &-0.306 &0.255 &-0.306 &-0.255 &30.482 &0.000 \\ 
-0.000 &-0.453 &0.255 &-0.011 &-0.255 &-0.011 &-0.000 &30.482 \\ 
\end{array}\right) $\\ 
\hline  
YW & $ \left( \begin{array}{cccccccc}74.949 &0.000 &-1.860 &-1.509 &-1.860 &1.509 &0.753 &0.000 \\ 
-0.000 &74.949 &-1.509 &-0.118 &1.509 &-0.118 &0.000 &-2.731 \\ 
-1.860 &-1.509 &74.949 &0.000 &0.753 &0.000 &-1.860 &1.509 \\ 
-1.509 &-0.118 &-0.000 &74.949 &0.000 &-2.731 &1.509 &-0.118 \\ 
-1.860 &1.509 &0.753 &0.000 &74.949 &0.000 &-1.860 &-1.509 \\ 
1.509 &-0.118 &-0.000 &-2.731 &-0.000 &74.949 &-1.509 &-0.118 \\ 
0.753 &0.000 &-1.860 &1.509 &-1.860 &-1.509 &74.949 &0.000 \\ 
-0.000 &-2.731 &1.509 &-0.118 &-1.509 &-0.118 &-0.000 &74.949 \\ 
\end{array}\right) $\\ 
\hline  
\end{tabular}  
}  
\end{table*}  